\documentstyle[12pt]{article}

\def\bfone{\relax{\rm 1\kern-.35em 1}}
\def\inbar{\vrule height1.5ex width.4pt depth0pt}
\def\IC{\relax\,\hbox{$\inbar\kern-.3em{\rm C}$}}
\def\ID{\relax{\rm I\kern-.18em D}}
\def\IF{\relax{\rm I\kern-.18em F}}
\def\IH{\relax{\rm I\kern-.18em H}}
\def\II{\relax{\rm I\kern-.17em I}}
\def\IN{\relax{\rm I\kern-.18em N}}
\def\IP{\relax{\rm I\kern-.18em P}}
\def\IQ{\relax\,\hbox{$\inbar\kern-.3em{\rm Q}$}}
\def\IR{\relax{\rm I\kern-.18em R}}
\font\cmss=cmss10 \font\cmsss=cmss10 at 7pt
\def\ZZ{\relax\ifmmode\mathchoice
{\hbox{\cmss Z\kern-.4em Z}}{\hbox{\cmss Z\kern-.4em Z}}
{\lower.9pt\hbox{\cmsss Z\kern-.4em Z}}
{\lower1.2pt\hbox{\cmsss Z\kern-.4em Z}}\else{\cmss Z\kern-.4em
Z}\fi}

 \def\c{\gamma}

\def\cF{{\cal F}} 
 
 \def\cK{{\cal K}}
\def\cL{{\cal L}} 
\def\cN{{\cal N}}

\def\tilde{\widetilde}
\def\bar{\overline}

\def\hat{\widehat}

\def\Coe#1.#2.{{#1\over #2}}
\def\coeff#1#2{\relax{\textstyle {#1 \over #2}}\displaystyle}
\def\coe#1.#2.{\relax{\textstyle {#1 \over #2}}\displaystyle}

\def\to{\rightarrow}
\def\notin{\hbox{{$\in$}\kern-.51em\hbox{/}}}

\def\IE{\relax{{\rm I\kern-.18em E}}}

\def\IGam{\relax{{\rm I}\kern-.18em \Gamma}}

\def\IA{\relax{\hbox{{\rm A}\kern-.82em {\rm A}}}}

\def\B{\Sigma}

\def\dop#1{{\rm d}\hskip -1pt #1}

\def\ee#1{{\rm e}^{#1}}
\def\trace{{\rm Tr}\hskip 1pt}
\def\ii{{\rm i}}

\def\hp{{\IH\IP}^{4(m+3)}}
\def\omep{\omega^{\scriptscriptstyle +}}
\def\omepind#1{\omega^{\scriptscriptstyle +\hskip 2pt #1}}
\def\omem{\omega^{\scriptscriptstyle -}}
\def\omemind#1{\omega^{\scriptscriptstyle -\hskip 2pt #1}}
\def\omepm{\omega^{\scriptscriptstyle \pm}}
\def\omepmind#1{\omega^{\scriptscriptstyle \pm\hskip 2pt #1}}
\def\Omep{\Omega^{\scriptscriptstyle +}}
\def\Omepind#1{\Omega^{\scriptscriptstyle +\hskip 2pt #1}}
\def\Omem{\Omega^{\scriptscriptstyle -}}
\def\Omemind#1{\Omega^{\scriptscriptstyle -\hskip 2pt #1}}
\def\Omepm{\Omega^{\scriptscriptstyle \pm}}
\def\Omepmind#1{\Omega^{\scriptscriptstyle \pm\hskip 2pt #1}}

\def\Jpind#1{J^{\scriptscriptstyle +\hskip 2pt #1}}

\def\Jmind#1{J^{\scriptscriptstyle -\hskip 2pt #1}}

\def\Jpmind#1{J^{\scriptscriptstyle \pm\hskip 2pt #1}}

\def\Jmpind#1{J^{\scriptscriptstyle \mp\hskip 2pt #1}}

\def\Qmind#1{Q^{\scriptscriptstyle -\hskip 2pt #1}}

\def\FFpind{F^{\scriptscriptstyle + }}
\def\FFmind{F^{\scriptscriptstyle -}}

\begin{document}
\begin{titlepage}
\hskip 0.1cm
\vbox{
\hbox{{\tt hep-th/9505123}}
\hbox{May, 1995}\hbox{\null} }
\hskip 2cm
\vbox{\hbox{SISSA 48/95/EP}\hbox{POLFIS-TH 06/95}
\hbox{CERN-TH 95/115}}
\hskip 2cm
\vbox{\hbox{IFUM 504/FT}\hbox{KUL-TF-95/15}
\hbox{UCLA/95/TEP/8}}
\vskip 0.1cm
\vfill
\begin{center}
{\LARGE R-Symmetry and the Topological Twist of \\
\vskip 2mm
N=2 Effective Supergravities of \\
\vskip 2mm
Heterotic Strings$^{\dagger}$}
\vskip 0.2cm
\vfill
{\large Marco Bill\'o$^1$, Riccardo D'Auria$^2$, Sergio Ferrara$^3$,\\
 \vskip 1.5mm Pietro Fr\'e$^{1,7}$, Paolo Soriani$^4$ and
Antoine Van Proeyen$^{5,6}$} \\
\vfill
\vskip 0.2cm
$^1$ International School for Advanced Studies (ISAS), via Beirut 2-4,
I-34100 Trieste\\
and Istituto Nazionale di Fisica Nucleare (INFN) - Sezione di Trieste\\
\vspace{6pt}
$^2$ Dipartimento di Fisica, Politecnico di Torino, c.so Duca degli
Abruzzi, Torino\\
and Istituto Nazionale di Fisica Nucleare (INFN) - Sezione di Torino\\
\vspace{6pt}
$^3$ CERN Theoretical Division, CH 1211 Geneva, Switzerland\\
and UCLA Physics Department, Los Angeles CA\\
\vspace{6pt}
$^4$ Dipartimento di Fisica, Universit\`a di Milano, via Celoria 16,
I-20133 Milano\\
and Istituto Nazionale di Fisica Nucleare (INFN) - Sezione di Milano\\
\vspace{6pt}
$^5$ Instituut voor Theoretische Fysica - Katholieke Universiteit Leuven
\\Celestijnenlaan 200D B--3001 Leuven, Belgium\\
\end{center}
\vskip 0.1cm
\begin{center}
{\bf Abstract}
\end{center}
\small
We discuss  R-symmetries in locally supersymmetric N=2 gauge
theories coupled to hypermultiplets which can be thought of as
effective theories of heterotic superstring models. In this
type of supergravities a suitable R-symmetry exists
and can
be used to topologically twist the theory: the vector multiplet
containing the dilaton-axion field has different R-charge
assignments with respect to the other vector multiplets.
Correspondingly a system of coupled instanton equations emerges,
mixing gravitational and  Yang--Mills instantons with triholomorphic
hyperinstantons and axion-instantons. For the tree-level
classical special manifolds $ST(n)=SU(1,1)/U(1)\times
SO(2,n)/(SO(2)$ $\times SO(n))$ R-symmetry with the specified
properties is a continuous symmetry, but for
the quantum corrected  manifolds ${\hat {ST}}(n)$ a discrete
R--group of electric--magnetic duality rotations is sufficient and
we argue that it exists.
\normalsize
\vskip 0.1cm
\vfill
\vfill \hrule width 3.cm
{\footnotesize
\noindent $^\dagger$ Partially supported by the EEC contract
sc1*-ct92-0789\\
\noindent $^6$ Onderzoeksleider, NFWO, Belgium.
E--mail : {\tt Antoine.VanProeyen@fys.kuleuven.ac.be}\\
\noindent $7$
E--mail: {\tt Fre@toux40.to.infn.it}  or {\tt Fre@tsmi19.sissa.it}}
\normalsize
\end{titlepage}
\section{Introduction}
\label{introduzione}
A large class of four
dimensional topological field theories can be obtained from the
topological twist of N=2 supergravity and N=2 matter theories [1--5].
The requirement that the twist should be well defined implies certain
additional properties on the scalar manifold geometries, besides those
imposed by N=2 supersymmetry, in order to obtain suitable ghost-number
charges and in order that the quaternionic vielbein be a Lorentz vector
after the twist. The needed properties pertain in particular to
those scalar manifolds which emerge, at the tree level, in the effective
theories of compactified superstrings.
Specifically they are:
\begin{enumerate}
\item[i)]{for the vector multiplet special manifold, an $R$-symmetry,
which is essential to redefine the ghost number of the fields after
the twist, and which, in the quantum case, is in general a discrete
symmetry};
\item[ii)]
for the hypermultiplet quaternionic manifold, an analogous
``Q-symmetry", which permits a consistent redefinition of
the Lorentz spin in the classical and quantum cases.
\end{enumerate}
Continuous R--symmetries are common features of the coset manifolds
which encode the local geometry of tree--level supergravity Lagrangians,
emerging as effective theories of N=2 heterotic superstrings.
The fact that the continuous symmetries present in the classical case
break to discrete symmetries is suggested, on physical grounds, by
the need to implement instanton corrections in the
effective lagrangian.
We will often refer
to the tree level (classical) theory
as to the ``microscopic" theory, as it is done in  paper
\cite{topf4d_2} by Seiberg and Witten, in contrast to the quantum or
effective  ``dynamical"
theory, where loop and instanton
corrections are taken into account and only the massless modes
are included.
\par
In this paper we are mainly concerned with the classical case, although we
give  indications of how our results can be generalized to the
quantum case.
Our main result,  namely the structure of the instanton
conditions that fix the topological symmetry, is independent of the detailed
form of the theory and simply follows from the existence of a discrete or
continuous R-symmetry
with the properties we shall require.  Hence the  form of these instanton
condition is universal and applies both to the classical and quantum case.
Specifically  it turns out that there are
four equations  describing the coupling of
four types of instantons:
\begin{equation}
\label{in1}
\begin{array}{rl}
\mbox{ i)} & \mbox{gravitational instanton}\\
\mbox{ ii)} & \mbox{gauge--instantons}\\
\mbox{ iii)} & \mbox{triholomorphic hyperinstantons}\\
\mbox{ iv)} & \mbox{H-monopoles}.
\end{array}
\end{equation}
Instanton equations of this type have already been discussed in
\cite{topftwist_1,topftwist_2,topf4d_8,topf4d_4};
the main difference is that in
\cite{topftwist_2,topf4d_8} the instanton conditions were only
the first three of eq.s (\ref{in1}).
The H-monopoles [8--12],
namely the instanton-like configurations
\begin{equation}
\label{in2}
\partial_{a} D\, =\,\epsilon_{abcd}\ee D H^{bcd}
\end{equation}
in the dilaton-axion sector were missing. In eq. (\ref{in2})
$D$ is the dilaton field and $H_{\mu\nu\rho}$ is the curl
of the antisymmetric axion
tensor $B_{\mu \nu}: \partial_{[\rho} B_{\mu\nu]}=H_{\mu\nu\rho}$.
The reason why they
were missing in
\cite{topftwist_2,topf4d_8} is the type of symmetry used there to define the
ghost number, namely an on-shell $R$-duality
based on the properties of the so-called minimal coupling.
The new type of gravitationally extended $R$-symmetry that we
present here is typically stringy in its origin and for
the classical moduli--spaces is an ordinary
off-shell
symmetry, which does not mix electric and magnetic states as the
$R$-duality
of the minimal case does. In the quantum--corrected effective
lagrangians $R$--symmetry reduces once again to an
$R$--duality, namely to a discrete
group of electric--magnetic duality rotations; yet the
preferred direction of the dilaton--axion field is
maintained also in the quantum case as it is necessary
on physical grounds. The new version of $R$--symmetry
discussed here
provides the solution to several conceptual problems at the same time.
\section{Outline and Philosophy}
\label{outline}
\setcounter{equation}{0}
Four dimensional Topological Field Theories,
which automatically select the appropriate
instanton conditions, are derived by topologically twisting
N=2, d=4 theories [1--5].
These latter include N=2 Yang--Mills theory, N=2 hypermultiplet
sigma models, N=2 supergravity, or else the coupling of all
such models together.
\par
In this paper we are concerned with the last case and with the
special features of the topological theory that emerge
when the parent N=2 matter coupled supergravity has the
structure of a low energy Lagrangian for an N=2 heterotic string theory.
\subsection{The geometry of vector multiplet and hypermultiplet
scalar manifolds}
\label{1geom}
As it is well known [13--27]
the Lagrangian and
the transformation rules of N=2 supergravity are completely
determined in terms of the following geometrical data:
\vskip 0.2cm
\par
1) The choice of a special K\"ahler manifold ${\cal SM}$ for
the vector multiplet scalars
\begin{equation}
{\rm dim}_{\IC} \,{\cal SM} = \, n + 1 \, {\stackrel{\rm def}{=}} \# \, {\rm
vector~multiplets}.
\label{ndefi}
\end{equation}
\par
2) The choice of a quaternionic manifold {\cal QM} for
the hypermultiplet scalars
\begin{equation}
{\rm dim}_{\IQ} {\cal QM}\, = {{1}\over{4}}
{\rm dim}_{\IR} {\cal QM}\, = \, m \, {\stackrel{\rm def}{=}}\,
\# \, {\rm hypermultiplets}.
\label{mdefi}
\end{equation}
\par
3) The choice of a gauge group ${\cal G}$ with:
\begin{equation}
\label{gaugegroup}
{\rm dim}_{\IR} {\cal G} \, \le \, n + 1 \,
\end{equation}
that generates special isometries of ${\cal SM}$ and should
have a triholomorphic action on the manifold ${\cal QM}$.
\vskip 0.2cm
In this paper, we are concerned with the following choices:
\begin{eqnarray}
\label{abc}
{\cal SM} &=& ST(n) \, {\stackrel{\rm def}{=}}\, {{SU(1,1)}\over
{U(1)}}\, \otimes {{SO(2,n)}\over{SO(2)\otimes SO(n)}}
\nonumber\\
{\cal QM}&=& HQ(m ) \, {\stackrel{\rm def}{=}}\,
{{SO(4,m)}\over{SO(4)\otimes SO(m)}}
\nonumber\\
{\cal G} & \subset &  SO(n)
\end{eqnarray}
where ${\cal G}$ is a $n$--dimensional subgroup of the $SO(n)$
appearing in the first equation above, such
that:
\begin{equation}
{\rm adjoint} \, {\cal G} \, = \, {\rm vector} \, SO(n) .
\label{adjoivector}
\end{equation}
\par
The structure given by eq. (\ref{abc}) is what one can
obtain by certain N=2 truncations of N=4 matter coupled supergravity
which, as it is well known, displays a unique coset structure:
\begin{equation}
{{SU(1,1)}\over
{U(1)}}\, \otimes {{SO(6,n+m)}\over{SO(6)\otimes SO(n+m)}}  \,
\supset  \,
 ST(n) \, \otimes \, HQ(m).
 \label{6contiene}
 \end{equation}
Other types of truncations can give different quaternionic coset
manifolds ${\cal QM}$ \cite{aggiunta}, for instance
$SU(2,m)/(SU(2) \times SU(m))$.
 Theories of type (\ref{abc}) originate, in particular cases, as
 tree-level low energy effective theories of the heterotic
 superstring compactified either on a $\ZZ_2$ orbifold
 of a six--torus $T^6 / \ZZ_2$ or on smooth manifolds
 of $SU(2)$ holonomy, like $ T_2 \otimes K3$
 \cite{skgsugra_16,ferpor,ferkouncos}, else when
 the superstring is compactified on abstract free fermion
 conformal field theories [30--33]
of type $(2,2)_{c=2} \, \oplus
(4,4)_{c=6}$ \cite{hmap_3}.
Although in the following we focus on the particular case where
${\cal QM}=HQ(m)$, our discussion on R-symmetry
is in fact concerned with the vector multiplet $ST(n)$ and applies
also when $HQ(m)$ is replaced by other manifolds.
Quantum corrections can change the geometry of $ST(n)$ or
 $HQ(m)$ in such a way that in the loop corrected Lagrangian
 they are replaced by new manifolds ${\hat {ST}}(n)$ or
 ${\hat {HQ}}(m)$, which are still respectively special
 K\"ahlerian and quaternionic, but
 which can, in principle, deviate from the round shape of coset
 manifolds.
 It is known that in rigid Yang--Mills theories coupled
 to matter the hypermultiplet metric (which is hyperk\"ahlerian)
 does not receive quantum corrections neither perturbatively, nor
 non--perturbatively \cite{sergius_1,ceretoine,lgvafa}.
 The same is true in N=2 supergravity theories derived from heterotic
 string theories: N=2 supersymmetry forbids a dilaton hypermultiplet
 mixing \cite{ceretoine,sergius_3,sergius_4}
since the dilaton is the scalar component of
 a vector multiplet. Hence in this case, while the scalar manifold is
replaced by $\hat{ST}(n)$, the quaternionic manifold $HQ(m)$ is
unmodified.
\par
The reverse is true (i.e. there are no quantum corrections to the vector
multiplet metric) for N=2 supergravities derived from type II strings
\cite{sergius_5}
\par
Generically continuous isometries break to discrete ones. This may be
a consequence both of ${\cal O} (\alpha^\prime )$ corrections  due to
the finite size of the string (discrete $t$--dualities generated by
non--perturbative world--sheet effects) and of non--perturbative
quantum effects due to space--time instantons (discrete Peccei Quinn
axion symmetries). Furthermore it can either happen that the
discrete quantum symmetries are just restrictions to special
values of the parameters of the classical continuous symmetries or
that they are entirely new ones. Usually the first situation occurs
when the local quantum geometry coincides with the local classical
geometry, namely when there are no corrections to the moduli space--metric
except for global identifications of points,  while the second situation occurs
when not only the global moduli geometry, but also the local one is quantum
corrected.
As we have stressed,
although ${\hat {ST}}(n)$ and ${\hat {HQ}}(m)$ may be quite different
manifolds from their tree level counterparts, they should still possess
an R-symmetry or a Q--symmetry so that the topological twist may be
defined.
Let us  then  discuss, on general grounds, the
problems related to the twist of  matter coupled N=2 supergravity.
\par
\subsection{The topological twist and the problem of ghost numbers}
\label{1toptwist}
In his first paper on topological field theories
\cite{topfgen_7}, Witten had shown how to derive a topological
reinterpretation of N=2 Yang--Mills theory in four--dimensions by
redefining the Euclidean Lorentz group:
\begin{equation}
SO(4)_{spin} \, =\, SU(2)_L \, \otimes \, SU(2)_R
\label{spingroup}
\end{equation}
in the following way:
\begin{equation}
SO(4)_{spin}^{\prime} \, =\, SU(2)_L \, \otimes \, SU(2)_R^\prime
\ ;\qquad SU(2)_R^\prime \, = \, {\rm diag} \left ( SU(2)_I \,
\otimes \, SU(2)_R \right )
\label{newspingroup}
\end{equation}
where $SU(2)_I$ is the automorphism group of N=2 supersymmetry.
In order to extend Witten's ideas to the case of an arbitrary
N=2 theory including gravity and hypermultiplets, four steps,
that were clarified in refs.
\cite{topftwist_1,topftwist_2}, are needed:
\begin{enumerate}
\item[i)] Systematic use of the BRST quantization, prior to the twist.
\item[ii)]
Identification of a gravitationally extended R-symmetry
that can be utilized to redefine the ghost--number in the
topological twist.
\item[iii)] Further modification of rule (\ref{newspingroup}) for the
redefinition of the Lorentz group that becomes:
\begin{equation}
SO(4)_{spin}^{\prime} \, =\, SU(2)_L^\prime \, \otimes \, SU(2)_R^\prime
\quad \quad \cases{
SU(2)_R^\prime \, = \, {\rm diag} \left ( SU(2)_I \,
\otimes \, SU(2)_R \right )\cr
SU(2)_L^\prime \, = \, {\rm diag} \left ( SU(2)_Q \,
\otimes \, SU(2)_L \right )\cr }
\label{newnewspingroup}
\end{equation}
Here $SU(2)_Q$ is a group whose action vanishes on all fields
except on those of the hypermultiplet sector, so that its role was not
perceived in Witten's original case.
\item[iv)] Redefinition of the supersymmetry ghost field {(\it topological
shift)}.
\end{enumerate}
Points { i)} and { iv)} of the above list do not impose
any restriction on the scalar
manifold geometry, so we do not discuss them further, although we shall
use the concept of topological shift in later sections.
(We refer the reader to  \cite{topftwist_1,topftwist_2} for further
details). Points ii) and iii), on the other hand have a bearing
on the geometry of ${\hat {ST}}(n)$ and ${\hat {HQ}}(m)$ and are
our main concerns.
\par
Topological field theories are cohomological theories of a suitable
BRST complex and as such they need a suitable ghost number $q_{gh}$
that, together with the form degree, defines the double
grading of the double elliptic complex.
In the topological twist, at the same time with the spin redefinition
(\ref{newnewspingroup}) one has a redefinition of the BRST charge
and of the ghost number, as follows:
\begin{eqnarray}
Q^{\prime}_{BRST} &=&  Q_{BRST} \, + \, \Qmind{0}_{BRST}
\nonumber\\
q_{gh}^{\prime} &=& q_{gh} \, + \, q_R .
\label{brstshift}
\end{eqnarray}
Here $Q_{BRST}$ is the old BRST charge that generates the BRST
transformations of the N=2 matter coupled supergravity and
$q_{gh}$ is the old ghost number associated with the BRST complex
generated by $Q_{BRST}$. We discuss now the shifts
$\Qmind{0}_{BRST}$ and $q_R$,
beginning with the former. The whole interest of the topological twist
is that $\Qmind{0}_{BRST}$ is just a component of
the Wick--rotated supersymmetry generators. It is defined
as follows.
\par
Writing the N=2 Majorana supercharges in the following bi-spinor
notation:
\begin{equation}
\IQ_A \, =\, \left ( \matrix { Q^{\alpha A} \cr
Q_{\dot\alpha A} } \right ) \quad \quad
\cases{\alpha=1,2\cr {\dot\alpha}=
\dot 1, \dot 2 , \cr}
\label{supercharges}
\end{equation}
so that a transformation of spinor parameter $\chi_A$ is
generated by:
\begin{equation}
\chi \, \cdot \, \IQ \, = \,
\chi_{\alpha A} \,Q^{\alpha A} \, + \, \chi^{
 \dot\alpha A} \, Q_{\dot\alpha A} ,
 \label{epsilontransform}
 \end{equation}
 we can perform the decomposition:
 \begin{equation}
Q_{\dot\alpha A}\, = \,
\epsilon_{\dot\alpha A} \,
\Qmind{0}_{SUSY} \, + \, \left (
\sigma_x \, \epsilon^{-1} \right )_{\dot\alpha A}
 \, \Qmind{x}_{SUSY}
 \label{susydecompo}
 \end{equation}
 and identify $\Qmind{0}_{SUSY}$ with the shift
 of the BRST charge introduced in eq. (\ref{brstshift}).
 It has spin zero as a BRST charge should have.
 In eq (\ref{susydecompo}) $\sigma_x$ are the standard Pauli matrices
 and $\epsilon_{AB}=-\epsilon_{BA}$, with
 $\epsilon_{12}=1$.
 Eq. (\ref{susydecompo})
makes sense because of the twist. Indeed, after $SU(2)_R$ has been
redefined as in eq.(\ref{newnewspingroup}) the isotopic doublet
index $A$ labeling the supersymmetry charges becomes an ordinary
dotted spinor index.
 \par
Let us now come to the discussion of the ghost number shift.
\subsection{$R$-symmetry in rigid N=2 theories}
\label{rigidrsym}
 The topological twist of a rigid N=2 supersymmetric Yang--Mills theory
 yields topological Yang--Mills theory, where the fields of the
 N=2 supermultiplet have the following reinterpretation:
 \begin{eqnarray}
 \mbox{gauge~boson} ~~A_\mu^\alpha &\rightarrow & {\rm phys.~field}
 \quad q_{gh}\, =
 \, 0 \nonumber\\
\mbox{left--handed gaugino} ~~\lambda^{\alpha A} &\rightarrow &
{\rm top.~ghost} \quad q_{gh}\, =
 \, 1 \nonumber\\
\mbox{right--handed gaugino} ~~\lambda_{\alpha^\star}^{A} &\rightarrow &
\mbox{top.~antighost} \quad q_{gh}\, =
 \, -1 \nonumber\\
\mbox{scalar} ~~Y^{I} &\rightarrow &
\mbox{ghost~for~ghosts} \quad q_{gh}\, =
 \, 2 \nonumber\\
\mbox{conjug.~scalar} ~~{\bar Y}^{I^\star} &\rightarrow &
\mbox{antighost~for~antighosts} \quad q_{gh}\, =
 \, -2
\label{ghostnumeri}
\end{eqnarray}
Hence, for consistency, the N=2 Yang--Mills theory should have,
prior to the twist, a global $U(1)$ symmetry with respect to which the
fields have charges identical with the ghost numbers they acquire after
the twist. In the minimal coupling case such a symmetry does indeed
exist and it is named R-symmetry (see for example \cite{fayet}).
\par
By minimal coupling we mean the situation where the {\it rigid
special geometry} \cite{topf4d_2,cerericca}
of the scalar manifold is defined by the following
generating function of quadratic type:
\begin{equation}
F(Y) \, = \, {\rm i} g_{IJ}^{(K)} \, Y^I \, Y^J
\label{quadrafunzia}
\end{equation}
where the $Y^I$
scalar fields are identified with
the rigid special coordinates and
$g_{IJ}^K$ is the constant Killing metric of the
gauge group ${\cal G}$. With such a choice, the K\"ahler metric
of the $Y$-scalar $\sigma$-model,
${\cal L}_{kin}^{scalars} \,= \,
g_{IJ^\star}(Y,{\bar Y}) \partial_\mu Y^I
\partial_\mu {\bar Y}^{J^\star}$ defined by:
\begin{eqnarray}
g_{IJ^\star}(Y,{\bar Y}) & = & \partial_I \,
\partial_{J^\star}
 \, {\cal K}_{rigid} (Y,{\bar Y}) \nonumber\\
 &=&{\rm i} \,  \partial_I \, \partial_{J^\star}
 \left ( Y^L \, \partial_{L^\star} {\bar F} \, - \,
 {\bar Y}^{L^\star} \, \partial_{L} F \, \right )\nonumber\\
 &= &2 {\rm Im}  \partial_I
 \,\partial_J \, F(Y)
 \label{rigidmetric}
 \end{eqnarray}
 takes the constant value:
 \begin{equation}
 g_{IJ^\star}\,= \, g^{(K)}_{IJ}
 \label{killingo}
 \end{equation}
 and the kinetic term
 ${\cal L}_{kin}^{vector} \,= \, {1\over 2\ii} [
{\cal N}_{IJ}({\bar Y})\,  {\FFpind}^I_{\mu\nu} \,
{\FFpind}^J_{\mu\nu}\, - \,
{\bar {\cal N}}_{IJ}(Y)\,  {\FFmind}^I_{\mu\nu}\,
{\FFmind}^J_{\mu\nu}] $ for the vectors, whose general definition
is provided by:
\begin{equation}
\bar {\cal N}_{IJ}(Y) \, =
\partial_{I}\partial_{J} F (Y)
\label{nmatrixrigid}
\end{equation}
is also of the standard form required for a renormalizable
gauge theory:
\begin{equation}
 {\cal L}_{kin}^{vector} \,= \,
-g^{(K)}_{IJ} \, \left ( {\FFpind}^I_{\mu\nu} \,
{\FFpind}^J_{\mu\nu} \, + \,
  {\FFmind}^I_{\mu\nu}\,
{\FFmind}^J_{\mu\nu} \right ) .
\label{kinovector}
\end{equation}
Minimal coupling corresponds, in the language of reference
\cite{topf4d_2}, to the {\it microscopic
gauge theory}. This theory has a scalar potential of the form:
\begin{equation}
V (Y,{\bar Y}) \,\propto\, g^{(K)}_{IJ} \,
f^{I}_{\  RS} \, f^J_{\ LM} \,
Y^R \, {\bar Y}^S \, Y^L \, {\bar Y}^M
\label{scalarepotenziale}
\end{equation}
where $f^{I}_{JK}$ are the structure constants of the group
${\cal G}$. The scalar potential has {\it flat directions},
namely it vanishes  for arbitrary values of
\begin{equation}
\left \{ \, Y \right \} \, \in \, \IC \, \otimes \, {\cal H}
\label{CSA}
\end{equation}
${\cal H}$ being the Cartan subalgebra of ${\cal G}$.
If we denote  by $Y^\alpha$ the scalar fields in the ${\cal H}$--subalgebra,
then $Y^\alpha$ are the moduli of the spontaneously broken gauge theory
that has ${\cal H}$ as unbroken gauge group and the components
$A^\alpha_\mu$ of the ${\cal H}$--connection
as massless gauge fields. The effective low energy
lagrangian for the massless modes is no longer a minimally
coupled N=2 gauge theory. Indeed its structure is determined
by a {\it rigid special geometry} encoded in a generating
function of the following form:
\begin{equation}
{\hat F} (Y^\alpha) \, = \, {\rm i} g_{\alpha\beta} Y^\alpha \,
Y^\beta
\, + \, \delta {\hat F} (Y^\alpha)
\label{instantonexpansion}
\end{equation}
where $g_{\alpha\beta}$ is the Cartan matrix of $\cal G$ and
$ \delta {\hat F} (Y^\alpha)$ accounts for the unique one--loop
correction  and for the infinite sum
of the instanton corrections \cite{sergius_2}.
In the topologically twisted theory the
deviation of the prepotential $F(Y)$ from the quadratic form
corresponds to perturbing the original minimal topological
lagrangian by means of all the available topological observables, namely
\begin{eqnarray}
S_{\min} &\longrightarrow & S_{\min} \, + \, \sum_{k} \, c(P_k) \,
\int _{{\cal M}_4} \, {\Phi}_{4, 4k-4}(P_k)\nonumber\\
 {\Phi}_{4, 4k-4}(P_k) & =&\mbox {4--form part of} ~ P_k ({\hat F} )
 \label{perturbazia}
 \end{eqnarray}
where $P_k(F)$
is any invariant polynomial of order $k$ of the gauge Lie algebra (i.e. a
characteristic class) and
where by ${\hat F}$ we have denoted the ghost--extended  field--strength
according to the standard rules of topological gauge theories in the
Baulieu--Singer set up
\cite{topfgen_5,topftwist_1,topftwist_2,fresoriabook}.
{}From this point of view the vacuum expectation values of the ghost fields
$Y^\alpha$ play the role of {\it ghost--charged}
topological coupling constants.
The continuous $R$-symmetry group $U(1)_R$ of the minimally
coupled theory is now broken, but either a discrete subgroup
${\hat G}_R \, \subset \, U(1)_R$ survives or a new discrete
R--group ${\hat G}_R^{quantum}$ replaces it.
\par
A similar phenomenon should occur in the gravitational case and this
is the matter of our study in the sequel.
 \subsection{$R$-symmetry in local N=2 theories and the moduli spaces of
 gravitational instantons}
 \label{rsimmetria}
 This being the situation in the rigid case, it is clear that,
 when N=2 supersymmetry is made local, $R$-symmetry should extend to
 a suitable symmetry of matter coupled supergravity.
 This problem was addressed in \cite{topftwist_2}, where it was
 shown that the minimally coupled local theory, which is also based
 on a quadratic generating function of the {\it local Special
 Geometry}:
 \begin{equation}
 F \, \left ( \, X^0 , X^\alpha \right )\, = \, i\left[\left ( X^0 \right )^2
 \, - \, \sum_{\alpha=1}^n \, \left ( X^\alpha \right )^2\right]
 \label{minimaleffe}                            ,
\end{equation}
 and which corresponds to the following choice for ${\cal SM}$:
 \begin{equation}
 {\cal SM} \, = \, {{SU(1,n)}\over{U(1)\otimes SU(n)}}
 \label{pippus}
 \end{equation}
 possesses an {\it $R$-duality}, namely an extension of $R$-symmetry
 that acts as a duality rotation on the graviphoton field strength,
 \begin{eqnarray}
 \delta F^{+ab}_{grav}&=&e^{i \theta} F^{+ab}_{grav}\nonumber\\
 \delta F^{-ab}_{grav}&=&e^{-i \theta} F^{-ab}_{grav}
 \end{eqnarray}
 mixing therefore electric and magnetic states.
 This result enabled the authors of \cite{topftwist_2} to discuss
 the topological twist in the case where the choice (\ref{pippus})
 is made.
 \par
 In this paper we show that the string inspired choice of
 eq. (\ref{abc}) yields another form of $R$-symmetry that allows
 the topological twist to be performed also in this case.
 Actually the new form of $R$-symmetry displays a new important feature
 that leads to the solution of a problem left open in the previous
 case.
 \par
 In the case (\ref{pippus}) all the vector fields, except the
 graviphoton, are physical since they have zero $R$-charge and hence
 zero ghost number after the twist. On the contrary, in this
 case, all the vector multiplet scalar fields
 are  ghost charged and hence unphysical.  The limit of pure topological
 gravity is obtained by setting $n=0$ in eq. (\ref{pippus}).
 This definition of 4D topological gravity \cite{topftwist_1} is
 correct but has one disadvantage that we briefly summarize.
 The topological observables of the theory
 \begin{equation}
 \int_{{\cal C}_2} \Phi_{(2,4n-2)} \, =
\int_{{\cal C}_2}\, \trace \, \left ( {\hat R} \, \wedge \,
 {\hat R} \, \wedge \, \dots \, \wedge {\hat R} \right )_{(2,4n-2)}
 \label{topolgraviobservables}
 \end{equation}
 (where ${\cal C}_2$ is a two cycle) have a ghost number which is always even
 being obtained from the trace of the product of an {\it even}
 number
 of (extended) curvature 2-forms (that this number should be even is
 a consequence of the self--duality of $R^{ab}$ in instanton
 backgrounds).
On the other hand
the moduli space of a typical  gravitational instanton (an ALE manifold)
has a moduli space with dimensionality \cite{gravinstant_1,%
gravinstant_3,gravinstant_4,gravinstant_5,momentmap_5}:
 \begin{equation}
 {\rm dim}_{\IC} \, {\cal M}_{moduli} (ALE)\, = \, 3 \, \tau
 \label{tretauproblem}
 \end{equation}
 $\tau$ being the Hirzebruch signature. It appears therefore
 difficult to saturate the sum rule
 \begin{equation}
 \sum_{i=1}^{n} gh^i \, = \, 3\, \tau
\label{sommarulla}
\end{equation}
needed for the non-vanishing of an $n$-point topological
correlator of local observables.
Notice, however, that it is possible to find nontrivial
topological correlation functions, satisfying the selection rule
(\ref{sommarulla}), between non local observables of the form
$\int_{{\cal C}_1}\Phi_{(1,4n-1)}$ for the topological
gravity with the Eguchi--Hanson instanton  \cite{nuovodamian}.
\par
 The origin of this problem is fairly evident to the string theorist
 and in particular to the string theorist who has experience
 with Calabi--Yau compactifications. Let us see why.
 The number $3\tau$ emerges
 as the number of deformations of the self-dual metrics on the
 ALE--manifold. To each self-dual harmonic 2-form one attaches
 a complex parameter (and hence 2 real parameters) for the deformations
 of the complex structure and a real parameter for the deformations
 of the K\"ahler structure, which sum to three parameters times
 the Hirzebruch signature. This counting, appropriate to pure
 gravity, is incomplete in the effective
 theory of superstrings where one has also the axion and the
 dilaton, besides the metric. An additional real modulus is associated
 with
 each selfdual 2-form for the deformations of the axion. This
 parameter can be used to complexify the complex structure
 deformations making the total dimension of moduli space
 $4 \tau$ rather than $3\tau$. Hence a sound 4-dimensional
 topological gravity should include also the dilaton and the axion,
 as suggested by the superstring. In the N=2 case these two fields
 are combined together into the complex field $S$, which is just the
 scalar field of an additional vector multiplet.
 Therefore we would like a situation where of the $n+1$ vector
 multiplets coupled to supergravity, $n$ have the ghost numbers
 displayed in eq. (\ref{ghostnumeri}), while one behaves in the
 reversed manner, namely:
\begin{eqnarray}
 \mbox{gauge~boson} ~~A_\mu^s &\rightarrow & {\rm ghost.~for~ghost}
 \quad q_{gh}\, =
 \, 2 \nonumber\\
\mbox{left--handed gaugino} ~~\lambda^{s A} &\rightarrow &
{\rm top.~antighost} \quad q_{gh}\, =
 \, -1 \nonumber\\
\mbox{right--handed gaugino} ~~\lambda_{s^\star}^{A} &\rightarrow &
\mbox{top.~ghost} \quad q_{gh}\, =
 \, 1 \nonumber\\
\mbox{scalar} ~~S &\rightarrow &
\mbox{\rm phys.~field} \quad q_{gh}\, =
 \, 0 \nonumber\\
\mbox{conjug.~scalar} ~~{\bar S} &\rightarrow &
\mbox{\rm phys.~field} \quad q_{gh}\, =
 \, 0 .
\label{ghostnumeriprimo}
\end{eqnarray}
This phenomenon is precisely what takes place in the new form
of $R$-duality, which is actually an $R$-symmetry, which applies
to the classical manifold $ST(n)$.
\par
The proof of this statement
is one of the main points of the present paper.
\par
In the quantum case we should require that the same $R$-charge assignments
(\ref{ghostnumeri}) and
(\ref{ghostnumeriprimo}) holds true. For this to be
true it suffices, as stressed in the introduction, that only a
(suitable) discrete R-symmetry survives.
\subsection{Gravi--Matter Coupled Instantons}
\label{gravimatter}
Provided the above restrictions on the scalar manifolds are
implemented one can describe in general terms the coupled
matter, gauge and gravitational instantons that arise from
the topological twist by means of the following equations:
\begin{eqnarray}
R^{-ab } \, - \, \sum_{u=1}^3 J_{u}^{-ab}
q^\star \hat \Omemind u   &=& 0\nonumber\\
\partial_a D \, - \, \epsilon_{abcd}\ee D H^{bcd}&=&0 \nonumber\\
{\cal F}^{-\,\alpha\,ab}\, - \, {{g}\over{2 \, \exp{D}}} \,
\sum_{u=1}^3 J^{-\, ab}_u {\cal P}_{\alpha}^{-\, u} &=& 0  \nonumber\\
{\cal D}_\mu q^P \, - \, \sum_{u=1}^{3} (j_u)_{\mu}^{~\nu}
\, {\cal D}_\nu q^Q \, ( J_u )_{Q}^{~P}& =& 0 .
\label{istantonequazioni}
\end{eqnarray}
In the above equations
$R^{- ab}$ is the antiselfdual part of the Riemann curvature
2--form ($a,b$ are indices in the tangent of the space time manifold),
$q^\star \hat \Omemind u$ denotes the pull--back,  via a
gauged--triholomorphic  map:
\begin{equation}
q \, : \, {M}_{space-time} \, \longrightarrow \,
HQ(m)
\label{triolomorfa}
\end{equation}
of the  ``gauged" 2--forms $\hat \Omemind u$
corresponding to one of the two quaternionic structures
of $HQ(m)$ (see appendices A and C)\footnote{Notice that
the instanton equation have the same expression
also in the quantum case
$\hat{ST(n)}, \hat {HQ(m)}$}.
${\cal P}_{\alpha}^{-\, u}$ are the corresponding
momentum map functions for the triholomorphic action of the gauge group
${\cal G}$ on ${HQ}(m)$. Furthermore $J_u^{-ab}$ is
nothing else but a basis of anti-selfdual matrices in $\IR^4$.
The second of equations (\ref{istantonequazioni}) describes the H--monopole
or {\it axion--dilaton instanton} first considered by Rey in
\cite{rey_1} and subsequently identified with the Regge--D'Auria
torsion instantons \cite{dauriaregge} and also with
the semi--wormholes of Callan et al \cite{callanstro}
according to the analysis of
\cite{gravinstant_1}. In the Rey formulation, that is the one
appearing here, the H--monopoles have vanishing stress--energy
tensor, so that they do not interfere with the gravitational
instanton conditions. The last of eq.s (\ref{istantonequazioni})
is the condition of triholomorphicity of the map (\ref{triolomorfa})
rewritten with covariant rather than with ordinary derivatives.
Such triholomorphic maps are the four--dimensional
$\sigma$--model instantons, or {\it hyperinstantons}
\cite{topftwist_2,topf4d_8}. Finally, in the same
way as the first of eq.s (\ref{istantonequazioni}) is the
deformation of the gravitational instanton equation due to
the presence of hyperinstantons, the third expresses the
modification of Yang--Mills instantons due to the same
cause. The space--time metric is no longer self--dual yet
the antiself--dual part of the curvature is just expressed in terms
of the hyperinstanton quaternionic forms. The same happens to
the antiself--dual part of the Yang--Mills field strength.
Deleting the first three of eq.s (\ref{istantonequazioni})
due the gravitational interactions one obtains the appropriate
generalization to any gauge--group and to any matter sector
of the so called monopole--equations considered by Witten
in \cite{topf4d_4}. That such equations were essentially contained
in the yield of the topological twist, as analysed in
\cite{topftwist_2},
was already pointed out in \cite{topf4d_8}.
The main novelty here
is the role played by the dilaton--axion sector that, as already
emphasized, should allow the calculation of non--vanishing
topological correlators between local observables
as intersection numbers in a moduli--space
that has now an overall complex structure.
\subsection{Topological gauge fixing as supersymmetric
\protect\\ backgrounds}
To find supersymmetric backgrounds of a supersymmetric theory,
one usually  looks for solutions of the eq.s:
\begin{equation}
\delta_{SUSY} \psi ~=~0
\label{fondamento}
\end{equation}
where $\psi$ is any fermion of the theory. In a generic $N=2$ theory which
includes
the dilaton there are four types of fermion, namely the gravitino, the
dilatino,
the gauginos and the hyperinos. Correspondingly there are four sets of
differential equations to be satisfied by the bosonic backgrounds. In
euclidean signature there exist non trivial solutions that preserve at most
half of the supersymmetries, namely those with supersymmetry parameter
of a given chirality. A set of solutions of these equations is
obtained
precisely by solving eq.s (\ref{istantonequazioni}). The reason for this is
that the instanton conditions (\ref{istantonequazioni}) are the
BRST--variations of the topological antighosts which in the untwisted
version of the theory coincide with the supersymmetry variations
of the fermions of a given chirality. The only difference is that in
eq.s (\ref{istantonequazioni}), in addition to the fermions we have set
to zero also those bosonic fields that have non--zero ghost number,
namely $Y^\alpha, A_\mu^{s} , A_\mu^{0}$.
\def\cds#1#2{#1\hskip -3pt\cdot\hskip -3pt #2} 
\def\cY{{\cal Y}}
\section{Special
geometry of the $ST(n)$ manifolds}
\label{specialerie}
\setcounter{equation}{0}
In this section we recall some general properties of special
K\"ahler manifolds \cite{skgsugra_3,skgsugra_8,skgsugra_2,%
skgsugra_4,skgmat_1,skgsugra_1,skgsugra_11},
then we focus on the manifolds
$ST(n)$ listed in eq. (\ref{abc}).
\par
Special geometry is the natural geometric structure which arises
in the coupling of N=2 four dimensional supergravity to vector
multiplets.
In particular, when
N=2 supergravity is regarded as an effective theory for the massless
modes of the compactified heterotic
string, the vector multiplets have a well
defined structure. Fixing their number to be $n+1$ we have that
$n$ of them
contain the ordinary gauge vectors:
\begin{equation} \label{so2n1}
(A^{\alpha}, \lambda^{\alpha A}, \lambda^{\bar\alpha}_A, Y^{\alpha}),
\hskip 0.3cm \alpha =1,\ldots , n
\end{equation}
and one:
\begin{equation}
\label{so2n2} (A^S, \lambda^{SA}, \lambda^{\bar S}_A, S)
\end{equation}
contains the dilaton-axion field:
\begin{eqnarray}
S &=&{\cal A}+ i\ee D \nonumber\\
\nabla_\sigma
{\cal A} &=&{\epsilon_{\sigma\mu\nu\rho}\over\sqrt{|g|}}
\ee{2D} H^{\mu\nu\rho} =
{\epsilon_{\sigma\mu\nu\rho}\over \sqrt{|g|}}
\ee{2D}\partial^{\mu}B^{\nu\rho} . \label{so2n3}
\end{eqnarray}
In eq.s (\ref{so2n1}) and (\ref{so2n2}) $A$ denotes the
gauge connection 1-form, $\lambda^A$ and $\lambda_A$ denote the
left-handed and the right-handed parts of the gauginos ($\gamma_5
\lambda^A = \lambda^A$, $\gamma_5 \lambda_A = -\lambda_A$) and
($S,Y^\alpha$)
are the complex scalar fields of the corresponding multiplets.
The N=2 supersymmetry imposes specific constraints on the
manifold spanned by the scalar fields. This manifold must be
a Hodge--K\"ahler manifold of restricted type, namely a special
K\"ahler manifold \cite{skgsugra_3,skgsugra_8,skgsugra_4,skgsugra_1}.
\par
In a generic $n+1$-dimensional
special K\"ahler manifold, the K\"ahler two form
can be expressed by the formula
\begin{equation} \label{so2n4} K={{\ii}\over {2\pi}} \partial\bar
\partial \log || W(S, Y)||^2= -{{\ii}\over{2\pi}}\partial\bar\partial
\log
||\Omega||^2 \end{equation}
where $W(S, Y)$ is a holomorphic section of
the Hodge line bundle ${\cal L}_{H} \stackrel{\pi}\rightarrow
 {\cal SM}$
and
\begin{equation} \label{so2n5} \Omega= (X^\Lambda, F_\Lambda)
\end{equation}
is a holomorphic section of ${\cal L}_H^2 \times {\cal
SP}$, where ${\cal SP} \stackrel{\pi^\prime}{\rightarrow} {\cal SM}$
 is a
flat, rank $2n+4$-vector bundle, with $SP(2n+4,\IR)$ structural group.
This amounts to say that the K\"ahler potential ${\cal K}$ has the
following expression:
\begin{equation} \label{so2nkpot} \cK = -\log
\left[ -\ii ({\bar X}, \bar F)^T \left( \begin{array}{cc} {0} & {\bfone
}\\-{\bfone } & {0}\end{array} \right) \hskip 3pt\left(\begin{array}{c}
X\\F\end{array}\right)\right] . \end{equation}
The symplectic index $\Lambda$ runs over $n+2$ values, and in the
cases related to string compactifications
it has the following labels:
$\{0,S, \alpha \}$ ($\alpha=1,\ldots n$), the index zero being
associated
to the gravitational multiplet.
\def\o#1#2{{#1\over#2}}
\par
In many bases, but not
necessarily in all bases, the symplectic section (\ref{so2n5}) can be
chosen in such a way that
\begin{equation}
\label{so2n5bis}
F_\Lambda={{\partial F(X)}\over{\partial X^\Lambda}} \end{equation}
where
$F(X)$ is a degree two homogeneous function of the $X^\Lambda$
coordinates, named
the prepotential.\par
Eq.s (\ref{so2n4}) and (\ref{so2nkpot}) implies that the Riemann
tensor for a generic
special manifold satisfies the following
identity :
\begin{equation}
R_{ij^*lk^*}= g_{i j^*}g_{l k^*}+ g_{ik^*}g_{l j^*}
-C_{ilp}C_{j^*k^*p^*}g^{pp^*}
\end{equation}
where $C_{ijk}=e^{\cal K}W_{ijk}$
are suitable sections of ${\cal L}_H^2 \times [T^{(1,0)}{\cal SM}]^3$.
These sections have a double physical interpretation.
In the N=2 effective lagrangians they play
the role
of fermionic anomalous magnetic moments , while, in
the associated N=1 theories (obtained from the N=2 ones
via the $h$--map \cite{hmap_2,hmap_3,hmap_1}), they can be interpreted
 as
Yukawa couplings.
\par
The elements of the symplectic structural group
$SP(2n+4,\IR)$, namely matrices with the following block structure
\begin{equation}
\label{so2nspmat}
\left(\begin{array}{cc}A & B\\C &
D\end{array}\right) \hskip 2cm \left\{\begin{array}{l}A^TC - C^T A =
0\\B^T D - D^T B = 0\\ A^T D - C^T B = {\bf 1}\end{array}\right.
\end{equation}
induce coordinate transformations on the scalar manifold
while acting, at the same time, as duality rotations on the symplectic
vector of electric and magnetic field strengths
\footnote{The $\cF_{ab}$
are the components along the space-time vierbeins of the field-strength
$\cF$ and $\cF^{\pm}_{ab}$ their (anti)selfdual
projections. The (anti)selfdual parts satisfy $\epsilon_{abcd}\cF^{\pm
cd}= \pm 2\ii\cF^{\pm}_{ab}$ and are defined by $\cF^{\pm ab} = {1\over
2}\left( \cF^{ab} \pm {\tilde\cF}^{ab}\right) $ where
${\tilde\cF}^{ab} = -{\ii\over
2}\epsilon^{abcd} \cF_{cd}$ is the dual tensor.}:
\begin{equation}
\label{so2n82} (\cF^{-\Lambda}_{ab}, G^{-}_{\Lambda ab}) \hskip 1cm
\mbox{where} \hskip 0.2cm G^{-}_{\Lambda ab}= -\ii {\delta\cL\over
\delta\cF^{-\Lambda}_{ab}},
\end{equation}
In the case the scalar manifold ${\cal SM}$
admits a continuous
or discrete isometry group $G_{iso}$, this group must be
suitably embedded into the duality group $SP(2n+4,\IR)$ and
the corresponding duality rotations, induced by the embedding,
leave
form invariant the system of Bianchi identities plus equation of motion
\cite{sugradual_1}.
\par
In this paper we are mainly concerned with the case ${\cal SM}=ST(n)$,
and, in the sequel, we focus our attention to its particular properties.
\par
The special K\"ahler manifold $ST(n)$ has been studied using different
parametrizations, corresponding to different embeddings of
the isometry
group $SL(2,\IR) \times SO(2,n)$ into the symplectic group
$SP(2n+4,\IR)$.
The first studied parametrization was based on a cubic type
prepotential $F(X) = \o{1}{X^0} X^S X^r X^t \eta_{rt}$, where
 $\eta_{rt}$
is the constant diagonal metric with signature $(+, -, \ldots, -)$
in a $n$-dimensional space \cite{skgsugra_2}.
In this parametrization  only an $SO(n-1)$ subgroup of
$SO(2,n)$ is linearly realized and it is possible to gauge only up to
$n-1$ vector multiplets. This means that, of the $n$ ordinary gauge vectors
sitting in the $n$ vector multiplets, only $n-1$ can be gauged.
\par
{}From a string
compactification point of view one does not expect this restriction:
it
should be possible to gauge all the $n$ vector multiplets containing
the
ordinary gauge vectors $A^\alpha$. This restriction motivated
the search for a second parametrization, where the $SO(n)$
subgroup is linearly realized. This parametrization is based on the
``square root" prepotential $F(X)=\sqrt{(X_0^2+X^2_1)X^\alpha
 X^\alpha}$ \cite{skgsugra_7}.
\par
However, in principle,
it should be possible to find a linear realization of the full
$SO(2,n)$ group, as it is predicted by the
tree level string symmetries.
In this case one can also gauge the graviphoton and the
gauge field associated to the dilaton multiplet.
This is explicitly realized in a recent work
\cite{ceretoine}, where the
new parametrization of the symplectic section is based on the
 following
embedding of the isometry group $SO(2,n)\times SL(2,\IR)$ into
$SP(2n+4,\IR)$.
\def\twomat#1#2#3#4{\left(\begin{array}{cc}#1& #2\\ #3 &
#4\end{array}\right)}
\begin{equation} \label{so2nemb} \begin{array}{ccc}
A\in SO(2,n) &\hookrightarrow &\left(\begin{array}{cc}A & {\bf 0}
\\ {\bf
0} & \eta A \eta^{-1} \end{array}\right)\in Sp(2n + 4,{\IR})\\ & & \\
\left(\begin{array}{cc}a & b\\c & d\end{array}\right) \in SL(2,
\IR)&\hookrightarrow & \twomat{a\bfone}{b\eta^{-1}}{c\eta}{d\bfone}\in
 Sp(2n + 4,\IR)\ , \end{array} \end{equation}
where $A^T\eta A=\eta$.
Notice that, in this embedding, the $SO(2,
n)$ transformations, when acting on the section $(\cF^{-\Lambda}_{ab},
G^{-}_{\Lambda ab})$, do not mix the $\cF$ with the $G$'s.
 Thus the true
duality transformations mixing the equations of motion and Bianchi
identities are generated by the embedding of the $SU(1,1)$ factor
 only,
so that the field $S$, that in our case parametrizes the coset $SU(1,
1)/U(1)$, plays a very different role from the $Y^{\alpha}$ fields.
\par
The explicit form of the symplectic section
corresponding to the
embedding  of eq. (\ref{so2nemb}) is:
\begin{eqnarray} \label{so2nssec}
(X^{\Lambda}, F_{\Lambda})&=&(X^{\Lambda}, S \eta_{\Lambda\Sigma}
X^{\Sigma})\nonumber\\ &&\nonumber\\ X^{\Lambda} &=&
\left(\begin{array}{c} 1/2\hskip 2pt (1 + Y^2) \\ \ii/2\hskip 2pt (1 -
Y^2)\\ Y^{\alpha}\end{array}\right).
\end{eqnarray}
In eq.
(\ref{so2nssec}) $Y^\alpha$ are
the Calabi--Visentini coordinates,
parametrizing the coset manifold
$SO(2,n)/SO(2)\times SO(n)$. The
pseudoorthogonal metric $\eta_{\Lambda\Sigma}$
has the signature $(+, +,
-, \ldots, -)$.\par Notice that, with the choice (\ref{so2nssec}),
 it is
not possible to describe $F_\Lambda$ as derivarivatives of any
prepotential. The K\"ahler potential for $ST(n)$ is obtained inserting in
eq. (\ref{so2nkpot}) the explicit form of the section
(\ref{so2nssec}), namely:
\begin{equation} \label{so2nkpotskn}
\cK=\cK_1(S, \bar S)+\cK_2(Y, \bar Y) = -\log \ii (\bar S - S) -
\log {\bar X}^T\eta X .
\end{equation}
{}From eq. (\ref{so2nkpotskn})
it easy to see that the K\"ahler metric
has the following block diagonal structure:
\begin{equation} \label{so2n5ter} \left(\begin{array}{cc}g_{S\bar S} & {\bf
0}\\ {\bf 0} & g_{\alpha \beta^*}\end{array}\right)\hskip 1.5cm
\left\{\begin{array}{l} g_{S\bar S} = \partial_S \partial_{\bar S}
\cK_1 =
{-1\over (\bar S - S)^2}\\ g_{\alpha\beta^*}(Y, \bar Y)=
\partial_{\alpha}\partial_{\beta^*} \cK_2 . \end{array}\right.
\end{equation}
The explicit expression of $g_{\alpha\bar\beta}(Y,\bar Y)$ is not
particularly relevant for our purposes. In the sequel, while
discussing
the instanton conditions, we will be interested only in its value at
$Y=0$ ($\bar Y=0$):
\begin{equation} \label{so2n6} g_{\alpha\beta^*}(Y=0) = 2
\delta_{\alpha\beta^*}. \end{equation}
The connection one form $Q$
of the line bundle $\cL_H$ is expressed in terms
of the K\"ahler potential as
\begin{equation}
Q^{(1,0)}+Q^{(0,1)}= \o{1}{2\ii}[ \partial_S \cK
\dop S + \partial_{\alpha}\cK \dop Y^{\alpha}]+ c.c .
\label{parapiglia}
\end{equation}
The explicit value of $Q^{(1,0)}$ at $Y=0$ is
\begin{equation}
\label{so2n7} Q^{(1,0)}(Y=0) =\o{1}{2} \o {\dop
S}{\bar S - S} .
\end{equation}
The anomalous magnetic moments-Yukawa
couplings sections $C_{ijk}$ ($i=S, \alpha$) have a very simple
expression
in the chosen coordinates:
\begin{equation}
\label{so2n6bis} C_{S\alpha\beta} = -\ee{\cK}
\delta_{\alpha\beta},
\end{equation}
all the other components being
zero.\par
In a general N=2 supergravity coupled to vector multiplets the
lagrangian for the vector bosons has a structure generalizing the
 rigid expression, namely
\begin{eqnarray} \label{so2n9}
\cL_{\rm kin}
&\propto & {1\over 2\ii} (\cN_{\Lambda\Sigma}\cF^{+\Lambda}_{ab}
\cF^{+\Sigma}_{ab} - {\bar\cN}_{\Lambda\Sigma}\cF^{-\Lambda}_{ab}
\cF^{-\Sigma}_{ab})\nonumber\\ &=& {1\over 2} ({\rm Im}
\cN_{\Lambda\Sigma} \cF^{\Lambda}_{ab} \cF^{\Sigma}_{ab}
-\ii {\rm
Re}\cN_{\Lambda\Sigma} {\cF}^{\Lambda}_{ab}
{\tilde\cF}^{+\Sigma}_{ab}) .
\end{eqnarray}
The general form of the
matrix $\cN_{\Lambda\Sigma}$ in the cases in which the prepotential
$F$
exists is given in \cite{skgsugra_8,skgsugra_2,skgsugra_1}.
Its further generalization,
including also
the cases where $F$ does not exists, has been found in
\cite{ceretoine}. In
our specific case, $\cN_{\Lambda\Sigma}$
is given by:
\begin{equation} \label{so2n10} \cN_{\Lambda\Sigma} = (S -
\bar S) {X_{\Lambda} {\bar X}_{\Sigma} + {\bar X}_{\Lambda} X_{\Sigma}
\over {\bar X}^T\eta X} + \bar S\eta_{\Lambda\Sigma}.
\end{equation}
In particular we have that ${\rm Re}\,\cN_{\Lambda\Sigma}$ =
${\rm Re}\, S
\eta_{\Lambda\Sigma}={\cal A} \eta_{\Lambda\Sigma}$. Moreover, at $Y=0$,
the only non-zero components of ${\rm Im}\cN_{\alpha\beta}$ are given by
\begin{equation}
{\rm
Im}\,\cN_{\alpha\beta}(Y=0) = {\rm Im}\, S\, \delta_{\alpha\beta}=
\exp{D}
\,\delta_{\alpha\beta} .
\end{equation}
Thus at $Y=0$ the kinetic term for the
ordinary gauge vectors
$A^\alpha$ reduces to $ \o{\rm
Im \,S}{g^2}\cF^{\alpha}_{ab} \cF^{\alpha}_{ab}$, where we have
explicitly taken into account the gauge coupling dependence, via
the usual redefinition $A^\alpha \to \o{1}{g} A^\alpha$.
This means that we can
reinterpret
$g_{\rm eff} = \o g{\sqrt{{\rm Im} S}}$ as the effective gauge
coupling.
\section{R-symmetry in N=2 Supergravity}
\label{rsimmetria2}
\setcounter{equation}{0}
In this section we give the general definition
of gravitationally extended $R$-symmetry.  Such a definition in
the continuous case pertains to  the $ST(n)$,
but in the discrete case
can be applied to much more general manifolds.
Furthermore it happens that in the classical $ST(n)$
case the continuous R--symmetry is an off--shell symmetry of the
action while in the quantum ${\hat {ST}}(n)$ case the
discrete R--symmetry acts in general as an electric--magnetic
duality rotation of the type of S--duality.
As stated in section \ref{rsimmetria}, the $R$-symmetry of
rigid N=2 gauge theories should have a natural extension
to the gravitationally
coupled case.
In principle, given a rigid supersymmetric theory, it is always possible
to define its coupling to supergravity, yielding a locally
supersymmetric theory. This does not mean that, starting with a complicated
``dynamical" N=2 (or N=1) lagrangian, it is an easy task to
define its gravitational extension. So we need some guidelines
to relate the R-symmetry of a rigid theory to the R-symmetry
of  a corresponding locally supersymmetric theory. The main points
to have in mind are the following ones:
\begin{itemize}
\item{The R-symmetry group $G_R$, whether continuous or discrete,
must act on the
symplectic sections $( X,\partial F )$ by means of symplectic
matrices:
\begin{equation}
\label{nuova}
\begin{array}{ccc}
\forall \, g \, \in \, G_R & \hookrightarrow & \left (\matrix{A(g) &
B(g) \cr C(g) & D(g) \cr } \right ) \, \in \, \Gamma_R \subset
SP(4+2n,\IR).
\end{array}
\end{equation} }
\item{The fields of the theory must have under $G_R$ well defined
 charges, so that
$G_R$ is either a $U_R(1)$ group if continuous or a cyclic group
$\ZZ_p$ if discrete.}
\item{By definition the left--handed and right--handed gravitinos must
have R--charges $q=\pm 1$, respectively}
\item{Under the $G_R$ action there must be, in the special manifold, a
preferred direction corresponding to the dilaton--axion multiplet
whose
R--charges are reversed with respect to those of all the other
multiplets. As emphasized, this is necessary, in order
for the topological twist to leave the axion--dilaton
field physical in the topological theory, contrary to
the other scalar partners of the vectors that become
ghosts for the ghosts}
\end{itemize}
 The last point of the above list is an independent assumption from
the previous three. In order to define a topological twist, the first
three properties are sufficient and are guaranteed by N=2 supersymmetry
any time the special manifold admits a symplectic isometry whose
associated K\"ahler rescaling factor is
$f_{2\vartheta}(z)=e^{2{\rm i}\theta}$ (see below for more
details). The third property characterizes the R--symmetry
(or R--duality) of those N=2 supergravities that have  an
axion--dilaton vector multiplet.
\par
For the classical coset manifolds
$ST(n)$  the appropriate
R-symmetry is continuous and it is easily
singled out: it is the
 $SO(2) \sim U(1)$ subgroup
of the isotropy group $SO(2) \times SO(n) \, \subset \, SO(2,n)$.
The coordinates that diagonalize the R--charges
are precisely the Calabi--Visentini coordinates discussed in the
previous section. In the flat limit they can be identified  with
the special coordinates of rigid special geometry. Hence such
gravitational R-symmetry is, as required, the supergravity
counterpart  of the R-symmetry considered in the rigid
theories.  Due to the direct product structure of
this classical manifold
the  preferred direction corresponding to the dilaton--axion field
 is explicitly singled out in the
$SU(1,1)/U(1)$ factor .
\par
Generically,
in the quantum case, the R-symmetry group $G_R$ is discrete.
Its action on the  quantum counterpart of the Calabi--Visentini
coordinates ${\hat Y}^\alpha$ must approach the action of a discrete
subgroup of the classical $U(1)_R$ in the same asymptotic region
where the local geometry of the quantum manifold ${\hat {ST}}(n)$
approaches that of $ST(n)$. This is  the large radius limit if we
think of
${\hat {ST}}(n)$ as of the moduli--space of some dynamical
Calabi--Yau threefold. To this effect recall that special K\"ahler
geometry is the moduli--space geometry of Calabi--Yau threefolds
and we can generically assume that any special manifold ${\cal SM}$
corresponds to some suitable threefold. Although the $G_R$
group is, in this sense, a subgroup of the classical $U_R(1)$
group, yet we should not expect that it is realized by a subgroup
of the symplectic matrices that realize $U(1)_R$ in the classical
case.  The different structure of the symplectic R--matrices is
precisely what allows a dramatically different form of the
special metric in the quantum and  classical case.
The need for this difference can be perceived {\it a priori} from
the request that the quantum R-symmetry matrix should be
symplectic integer valued. As we are going to see this is possible
only for $\ZZ_4$ subgroups of $U(1)_R$ in the original
symplectic embedding. Hence  the different
$\ZZ_p$ R--symmetries appearing  in rigid  quantum  theories
should have different  symplectic embeddings in the gravitational case.
\par
Let us now give the general properties of the gravitationally extended
$R$-sym\-me\-try, postponing to section \ref{rsimmetria3} the treatment of
the specific case $ST(n)$.
\par
\subsection{The general form of R-symmetry in supergravity }
 R-symmetry is either a $U(1)$ symmetry or
 a discrete $\ZZ_{p}$ symmetry.
Thus, if $R$-symmetry acts diagonally with charge $q_R$ on a field $\phi$,
this means  that $\phi\rightarrow \ee{q_R\ii \vartheta}
\phi$, $\vartheta\in [0,2\pi]$ in the continuous case.
In the discrete case only the values
$\vartheta=\frac{2\pi}{p}l$,  $l=0,1\ldots p-1$ are allowed
and in particular
the generator of the $\ZZ_{p}$ group acts as  $\phi\rightarrow
 R\phi =
\ee{q_R\frac{2\pi\ii}{p}}\phi$.
\par
By definition $R$-symmetry acts diagonally with charge $+1$ ($-1$)
on the left-(right)-handed gravitinos (in the same way as
it acts on the supersymmetry parameters in the rigid case):
\begin{equation}
\label{so2n11}
\begin{array}{c}
\psi_A\rightarrow\ee{\ii\vartheta}\psi_A\\
\psi^A \rightarrow\ee{-\ii\vartheta}\psi^A
\end{array}
\hskip 1cm \mbox{i.e.}\hskip 1cm
\begin{array}{l}
q_L(\psi_A) = 1\\
q_R(\psi^A) = -1 .
\end{array}
\end{equation}
$R$-symmetry
generates isometries  $z^i\rightarrow (R_{2\vartheta}z)^i$
\footnote{As in the rigid case,
the action of the $R$-symmetry group on the gravitinos, and more generally
on the fermions, {\it doubly covers} its action on the bosonic fields.
This property will become evident in eq. (\ref{ftheta}); it
explains the chosen notation $(M_{2\vartheta})^i_l$ for the matrix expressing
the $R$-action on the tangent bundle $T^{(1,0)}{\cal SM}$.}
of the scalar metric  $g_{ij^*}$ and it is embedded into
$Sp(2n+4,\IR)$ by means of  a symplectic matrix:
\begin{equation}
\label{Mmatrix}
 M_{2\vartheta} =
\left(\begin{array}{cc}  a_{2\vartheta} & b_{2\vartheta}
\\c_{2\vartheta} &
d_{2\vartheta}
\end{array}\right)   \, \in \, Sp(2n+4,\IR).
\end{equation}
\par\noindent
As we have already pointed out it turns out that
in the classical case of $ST(n)$ manifolds R--symmetry
does not mix the Bianchi identities with the field equations
since the matrix (\ref{nuova}) happens to be block diagonal:
$b_{2\theta}=c_{2\theta}$.
In the quantum case, instead, this is in general not true.
There is a symplectic action on the section
 $(X^\Lambda,F_\Lambda)$
induced by $z^i\rightarrow (R_{2\vartheta}z)^i$:
\begin{equation}
\label{fattorello}
(X,F)\rightarrow f_{2\vartheta}(z^i)  {M}_{2\vartheta} \cdot (X,F)
\end{equation}
where the K\"ahler compensating factor $f_{2\vartheta}(z^i)$ depends in
general both on the transformation parameter $\vartheta$ and on the
base--point $z$. By definition this compensating factor is the same
that appears in the transformation of the gravitino field $\psi_A \,
\to \, \exp [f_{2\vartheta}(z^i)/2] \, \psi_A$. Since  we have
imposed  that the transformation of the gravitino field should be as
in (\ref{so2n11}) it follows that the R-symmetry transformation
must be such as to satisfy eq.(\ref{fattorello}) with a suitable matrix
(\ref{Mmatrix}) and with a compensating K\"ahler factor
of the following specific form:
\begin{equation}
\label{ftheta}
f_{2\vartheta}(z^i) = \ee{2\ii\vartheta} .
\end{equation}
Condition $(\ref{ftheta})$ is a crucial constraint on the form of
$R$-symmetry.\par\noindent
The action of the R-symmetry on the matrix ${\cal N}$ is determined
by the form of the matrix $M_{2\vartheta}$ (see \cite{ceretoine}):
\begin{equation}
\label{ntransf}
{\cal N} \rightarrow (c_{2\vartheta} + d_{2\vartheta}{\cal N})
(a_{2\vartheta} + b_{2\vartheta}{\cal N} )^{-1}
\end{equation}
The supersymmetry transformation rules are encoded in the rheonomic
para\-me\-tri\-zations of the curvatures, summarized in Appendix C.
For instance the supersymmetry transformations of the scalar fields
are given by
\begin{equation}
\label{deltazi}
\nabla z^i = \nabla_a z^i V^a + {\bar\lambda}^{i A} \psi_A
\hskip 1cm \Rightarrow \hskip 1cm
\delta_{\epsilon} z^i = {\bar\lambda}^{i A} \epsilon_A .
\end{equation}
Let us denote by
$J$ the Jacobian matrix
of the transformations
\begin{equation}
(J_{2\vartheta})^i_l =
\frac{\partial (R_{2\vartheta} z)^i}
{\partial z^l} .
\end{equation}
If we now act on the scalars $z^i$ by an R-transformation we
conclude that,
using eq.s (\ref{deltazi},\ref{so2n11})
\begin{equation}
\label{deltalambdai}
\nabla z^i \rightarrow (J_{2\vartheta})^i_j \nabla z^j
\hskip 1cm \Rightarrow \hskip 1cm
\lambda^{i A} \rightarrow \ee{-\ii\vartheta}
(J_{2\vartheta})^i_j \lambda^{j A}
\end{equation}
Analogous considerations can be done for the hyperinos (cfr. eq.
(\ref{hy1})).\par\noindent
$\bullet$ The supersymmetry transformation of the
gravitino field are encoded in eq.s
(\ref{3.6b}, \ref{3.6c}, \ref{3.10a}, \ref{3.10b})
and in their gauged counterparts (\ref{bla1}, \ref{bla2}).
Requiring consistency  with eq. (\ref{so2n11})
determines the $R$-charges of the various terms in the right hand side.\par
{\it i)} The terms like $A^A_{\hskip 3pt B|b}$ that contain
bilinear in the fermions are neutral (cfr. eq.s (\ref{3.16a},\ref{3.16b}))
\par
{\it ii)} The R-symmetry acts diagonally on
terms $T^\pm_{ab}$.
These terms must have charge $q_R(T^\pm_{ab})=\pm 2$.
Notice that $T^\pm_{ab}$ can be
expressed by the following symplectic invariants (see eq. (\ref{c20}))
\begin{equation}
\label{tmunu}
T^-_{ab} \propto \ee{\frac{\cK}{2}} ({\bar X}^\Lambda,{\bar F}_\Lambda) \cdot
\left(\begin{array}{c}{\hat\cF}^{\Lambda -}_{ab}\\
\cN_{\Lambda\Sigma} {\hat\cF}^{\Sigma -}_{ab}\end{array}\right) ,
\end{equation}
where ${\hat\cF}^{\Lambda -}_{ab}$ $=\cF^{\Lambda -}_{ab} +$ $ \frac{1}{8}
\nabla_{i^*}{\bar f}^\Lambda_{j^*}{\bar\lambda}^{i^*}_A\gamma_{ab}
\lambda^{j^*}_B \epsilon^{AB}$.
Under an $R$ transformation the symplectic product appearing in eq.
(\ref{tmunu}) is left invariant up to the overall
(antiholomorphic) factor coming from eq. (\ref{fattorello}),
namely
\begin{equation}
\label{tm1}
T^-_{ab}\rightarrow \bar f_{2\vartheta}({\bar z}^{i^*})
T^-_{ab} .
\end{equation}
Since the R-symmetry act diagonally on $T^-_{ab}$ and
$q_R(T^-_{ab})=-2$, we necessarily have
\begin{equation}
\label{tm2}
T^-_{ab}\rightarrow
\ee{-2\ii\vartheta} T^-_{ab}.
\end{equation}
Eq.s (\ref{tm1}) and (\ref{tm2}) are consistent with eq. (\ref{ftheta}).
\par\noindent
$\bullet$ Let us consider the supersymmetry transformations
of the gauginos, encoded in eq.s (\ref{3.13a}, \ref{3.13b}) [and
their gauged counterpart (\ref{bla3},\ref{bla4})].
We impose that the Jacobian matrix is covariantly
constant, $\nabla (J_{2\vartheta})^i_j = 0$.
It then follows that the curvature $\nabla \lambda^{iA}$
transforms as $\lambda^{iA}$, that is as in eq. (\ref{deltalambdai}).
We can in this way verify that the R transformations of $G^{- i^*}_{ab}$
(and its complex conjugate) transform consistently with the gaugino
transformation.
\par
The terms $\cY^{i^*}_{AB}$
are proportional to the Yukawa couplings $C_{ijk}$.
These latter can be written in terms of a symplectic product:
\begin{equation}
\label{cijk}
\begin{array}{l}
C_{ijk} = (f_i,h_i)\cdot \nabla_j
\left(\begin{array}{c}f_k\\h_k\end{array}\right)  \\
(f^\Lambda_i,h_{\Lambda i}) \equiv \ee{\cK/2}\nabla_i (X^\Lambda,F_\Lambda).
\end{array}
\end{equation}
Their $R$-transformation is therefore\footnote{Indeed the section
$(f_i,h_i)$ transforms
into $\ee{2\ii\vartheta}{M}_{2\vartheta} ((J^{-1}_{2\vartheta})_i^{\hskip
3pt l}
f_l, (J^{-1}_{2\vartheta})_i^{\hskip 3pt l} h_l)$.
Then eq. (\ref{cijktransf}) follows.
Notice that this transformations is the appropriate one
for a section of $\cL^2_H\times [T^{(1,0)}{\cal SM}]^3$, that is
the correct interpretation
of the $C_{ijk}$'s.}:
\begin{equation}
\label{cijktransf}
C_{ijk}\rightarrow \ee{4\ii\vartheta} (J^{-1}_{2\vartheta})_i^{\hskip 3pt l}
(J^{-1}_{2\vartheta})_j^{\hskip 3pt m} (J^{-1}_{2\vartheta})_k^{\hskip 3pt n}
C_{lmn}.
\end{equation}
Utilizing eq. (\ref{cijktransf}) in eq. (\ref{3.13b})
one can check that the transformation of
$\cY^{i^*}_{AB} = g^{i^*j} C_{jlm} {\bar\lambda}^{l C} \lambda^{m D}
\epsilon_{AC}\epsilon_{BD}$
is consistent with the transformation of the left hand side.
\par
As can be easily verified, all the terms due to the gauging of the
composite connections transform in the correct
way to ensure the consistency of the R-transformations
[see eq.s (\ref{bla0}--\ref{c90})].
\vskip 0.1cm\par\noindent
$\bullet$ {\it Summarizing:}\par
The $R$-symmetry must act holomorphically on the scalar fields, $z^i
\rightarrow (R_{2\vartheta}\,z)^i(z)$, being an isometry.
Moreover the matrix $(J_{2\vartheta})^i_j$ has to be
{\it covariantly constant}:
$\nabla (J_{2\vartheta})^i_j = 0$.
The $R$-transformation of parameter $\vartheta$ on the scalar fields must
induce the transformation
$(X,F) \rightarrow \ee{2\ii\vartheta}  M_{2\vartheta} (X,F)$, where
$M_{2\vartheta}$ is of the form (\ref{Mmatrix}).
In the topological twist, the ghost numbers are redefined as in
eq. (\ref{brstshift})
by adding the $R$-charges.
\vskip 0.1cm\par\noindent
$\bullet$ {\it The dilaton--axion direction in the discrete case:}\par
In the classical case of the $ST(n)$ manifolds the existence of
a preferred direction is obvious from the definition of the
manifolds and R--symmetry singles it out in the way discussed
in the next section. Let us see how the dilaton--axion
direction can be singled out by the discrete R--symmetry
of the quantum manifolds ${\hat {ST}}(n)$.  Let $G_R=\ZZ_p$
and let $\alpha=e^{2\pi{\rm i}/p}$ be a $p$--th root of the
unity.  In the space of the scalar fields $z^{i}$ there always
will be a coordinate basis $\{ u^{i} \}$ $(i=1,\dots \, n+1)$
that diagonalizes the action of $R_{2\vartheta}$ so that:
\begin{equation}
R_{2\vartheta} \, u^{i} ~=~\alpha^{ q_i} \, u^{i} \quad \quad
q_i = 0,1, \dots , p-1 \, \mbox{mod} \, p
\label{drago_1}
\end{equation}
The $n+1$ integers $q_i$ (defined modulo $p$) are the
R--symmetry charges of the scalar fields $u_i$.
At the same time a generic $Sp(4+2n,\IR)$ matrix
has eigenvalues:
\begin{equation}
\left ( \lambda_0, \lambda_1, \dots , \lambda_{n+1}, {{1}\over{\lambda_0}},
{{1}\over{\lambda_1}},
\dots , {{1}\over{\lambda_{n+1}}} \right )
\label{autovalores}
\end{equation}
The R--symmetry symplectic matrix $M_{2\vartheta}$ of
eq. (\ref{Mmatrix}), being the generator of a cyclic group
$\ZZ_p$, has eigenvalues:
\begin{equation}
\lambda_0 = \alpha^{k_0}, \quad \lambda_1 = \alpha^{k_1}, \quad
\dots ,\ \lambda_{n+1} = \alpha^{k_{n+1}}\ ,
\label{specialmente}
\end{equation}
where $(k_0 , k_1 , \dots \, k_{n+1} )$ is a new set of
$n+2$ integers defined modulo $p$. These numbers are
the R--symmetry charges of the electric--magnetic field strengths
\begin{equation}
F_{\mu\nu}^0+{\rm i} \, G_{\mu\nu}^0, \quad F_{\mu\nu}^1
+ {\rm i} \, G_{\mu\nu}^1 ,\quad \dots \,\quad F_{\mu\nu}^{n+1}
+ {\rm i} \, G_{\mu\nu}^{n+1}\ ,
\end{equation}
their negatives, as
follows eq. (\ref{autovalores}), being the charges
of the complex conjugate combinations
$F_{\mu\nu} - {\rm i} G_{\mu\nu}$.
Since what is really
relevant in the topological twist are the differences of
ghost numbers (not their absolute values), the interpretation
of the scalars $u^{i}$ $(i=1, \dots , n)$ as ghost for ghosts and of
the corresponding vector fields as physical gauge fields
requires that
\begin{equation}
q_i ~=~k_i \, + \, 2 \quad\quad  i=1, \dots , n
\label{shifto}
\end{equation}
On the other hand, if the vector partner of the
axion--dilaton field has to be a ghost for ghosts,
the $S$--field itself being physical, we must
have:
\begin{equation}
k_{n+1} = q_{n+1} + 2
\label{inversamente}
\end{equation}
In eq. (\ref{inversamente}) we have conventionally identified
\begin{equation}
S=u^{n+1}
\end{equation}
Finally the R--symmetry charge $k_0$ of the last vector field--strength
$F^0_{\mu\nu}$
is determined by the already established transformation eq. (\ref{tm2})
of the graviphoton combination (\ref{tmunu})\footnote{
{\it [Note added in proofs]}
\par
In \cite{monodrcy} an explicit example is
provided of quantum R--symmetry based on the local N=2 $SU(2)$
gauge theory associated with the Calabi--Yau manifold
$WCP_4(8;2,2,2,1,1)$ of Hodge numbers $(h_{11}=2, h_{21}=86)$ that
has been considered by Vafa and Kachru in \cite{VafaKach} as an example of
heterotic/type II duality. }
\par
Summarizing, this situation is similar to that occurring in the
topological Landau--Ginzburg models \cite{lgvafa} where the
physical scalar fields $X^{i}$ have a non--zero R--symmetry
charge equal to their homogeneity weight
\cite{billusfre,fresoriabook} in the superpotential ${\cal W}(X)$.
After topological twist they acquire a non zero (fractional)
ghost--number that however differs from the ghost--number
of the fermions by the correct integer amount.
\footnote{{\it [Note added in proofs]} In a very recent
paper \cite{DJTTVd}
it has been proposed a new interpretation
of the D=2 topological Landau--Ginzburg models, based
on BRST, anti--BRST symmetry  where,
notwithstanding the fractional R--charges, the ghost numbers
become integer. It would be interesting to inquiry whether
such analysis can be extended to the case of D=4 vector
multiplets}
\subsection{$R$--symmetry in the $ST(n)$ case}
\label{rsimmetria3}
In the case of the microscopic lagrangian
the special K\"ahler manifold of the scalars is a
$ST(n)$ manifold.
The action of  $R$-symmetry is extremely simple.
As already stated in section 2, see eq. (\ref{ghostnumeriprimo}),
the $S$ field
has to be neutral, while the $Y^\alpha$ fields have $R$-charge 2:
\begin{equation}
\label{sknrsim1}
\left\{\begin{array}{l} S \rightarrow S\\ Y^\alpha \rightarrow
\ee{2\ii\vartheta} Y^\alpha \end{array}\right.
\hskip 0.5cm \Rightarrow (J_{2\vartheta})^i_j =
\left(\begin{array}{cc}1 & 0\\ 0 & \ee{2\ii\vartheta}\delta^\alpha_\beta
\end{array}\right) .
\end{equation}
Using the factorized form eq. (\ref{so2n5ter})
of the metric,
it is immediate to check
that the matrix $J_{2\vartheta}$ is covariantly
constant.\par
Utilizing the explicit form eq. (\ref{so2nssec}) of the symplectic section,
eq. (\ref{sknrsim1}) induces the transformation:
\begin{equation}
\label{sknrsim2}
\begin{array}{l}
\left(\begin{array}{c}X\\F\end{array}\right) \rightarrow \ee{2\ii\vartheta}
\left(\begin{array}{cc}{ m}_{2\vartheta} & 0\\ 0 &
({ m}^T_{2 \vartheta} )^{-1}\end{array}\right)
\left(\begin{array}{c}X\\F\end{array}\right)\\ \null \\
{ m}_{2\vartheta} =
\left(\begin{array}{ccc}\cos 2\vartheta & -\sin 2\vartheta & 0\\
\sin 2\vartheta & \cos 2\vartheta & 0\\
0 & 0 & \bfone_{n\times n}\end{array}\right)\in SO(2,n) .\end{array}
\end{equation}
We see that the crucial condition (\ref{ftheta}) is met. Furthermore note
that in this classical case $b_{2\vartheta}=c_{2\vartheta}=0$,
the matrix (\ref{Mmatrix}) is completely diagonal
and it has the required eigenvalues $(\ee{\ii\theta},\ee{-\ii\theta},
1,\ldots,1)$.\par
At this point we need no more checks; the $R$-symmetry defined by
eq. (\ref{sknrsim1}) is a true symmetry of the lagrangian and satisfies
all the expected properties.
The gauge fields $A^\alpha$ do not transform, while the $A^0,A^S$ gauge
fields undergo an $SO(2)$ rotation:
\begin{equation}
\label{sknrsim6}
\left\{\begin{array}{ccl}
\left(\begin{array}{c}A^0\\ A^S\end{array}\right) & \rightarrow
& \twomat{\cos 2\vartheta}{-\sin 2 \vartheta}{sin 2\vartheta}{\cos 2
\vartheta} \left(\begin{array}{c}A^0\\ A^S\end{array}\right) \\
\null & \null \\
A^\alpha & \rightarrow A^\alpha .
\end{array}\right.
\end{equation}
Notice that from eq.s (\ref{sknrsim2}), (\ref{sknrsim6})
and from the explicit form of the
embedding (\ref{so2nemb})
we easily check that the R-symmetry for the $ST(n)$
case is nothing else but the $SO(2)\sim U(1)$ subgroup
of the isometries appearing in the denominator of the
coset $SO(2,n)/SO(2)\times SO(n)$.
\par
At the quantum level the $R$--symmetries should act
on the symplectic sections as a symplectic matrix  belonging
to $Sp(2n + 4, \ZZ)$.  Consider then the intersection
of the continuous $R$ symmetry of eq.s (\ref{sknrsim1},\ref{sknrsim2})
with  $Sp(2n + 4, \ZZ)$: the result is a $\ZZ_4$ R-symmetry
generated by the matrix  $M_{2\vartheta}$ with $\vartheta=\pi /4$,
where:
\begin{equation}
\label{discreta}
m_{\pi /2}=\left(\begin{array}{ccc}0 & -1 & 0\\
1 & 0 & 0\\
0 & 0 & \bfone_{n\times n}\end{array}\right)\in SO(2,n; \ZZ) .
\end{equation}
As already observed, in a generic case, after the quantum
corrections are implemented, the discrete R-symmetry
$\ZZ_p$ is a subgroup of $U(1)_R$ as far as the action
on the moduli at large values is concerned, but it is
implemented by $Sp(4+2n,\ZZ)$ matrices that are
not the restriction to discrete value of theta of the
matrix $M_{2\vartheta}$ defined in eq.s (\ref{sknrsim2}).
In the one modulus case where,  according to the
analysis by Seiberg--Witten the rigid R-symmetry is $\ZZ_4$,
there is the possibility of maintaining the classical form of
the matrix $M_{2\vartheta}$ also at the quantum level and
in the case of local supersymmetry. This seems to be
a peculiarity of the one--modulus N=2 gauge theory.
\par
To conclude, in table 1 we give
the $R$-symmetry charge assignments for the
fundamental fields of the $ST(n)$ case together with the spin and
R-symmetry assignments for the hyperini and
for quaternionic vielbein $u$,
which will be properly defined in appendix A.
Notice that in this table,
concerning the quaternionic sector, we have explicitly
splitted the $SO(4)$ index $a$ (see appendix A for details)
into the $SU(2)_I \times SU(2)_Q$
indices $(A, \bar A)$ so that $u^{a t} \equiv u^{A~~t}_{~\bar A}$.
This splitting is fundamental, in order to redefine
correctly the Lorentz group for the twist, so that, after the
twist prescription \label{} is performed,
the quaternionic vielbein become a Lorentz vector. This
is consistent with the fact that $u$ appear in the
topological variation of $\zeta^{\bar A t}$, which acquires spin 1
after the twist. But we are going to analyse these
problems in the following section.
\begin{table}
\begin{center}
\caption{\sl Spin--R charges assignments}
\begin{tabular}{|c||c||c||c||c||c||c|}
\hline
Field   & $SU(2)_L$  & $SU(2)_R$  & $SU(2)_I$ & $SU(2)_Q$ & $R$ &
$gh^\prime$
\\
\hline
$V_a^\mu$ & $1/2$ & $1/2$ & $0$ & $0$ & $0$ & $0$  \\
\hline
$\psi_{\mu A}$ &$1/2$ & $0$ & $1/2$ & $0$ & $1$ & $1$ \\
\hline
$\psi_\mu^A$ & $0$ & $1/2$ & $1/2$ & $0$ &$-1$ &$-1$ \\
\hline
$A_\mu^0 +iA_\mu^S$ & $0$ &$0$ &$0$ &$0$ &$2$ & $2$  \\
\hline
$A_\mu^0 -iA_\mu^S$ & $0$ &$0$ &$0$ &$0$ &$-2$ & $-2$  \\
\hline
$A_\mu^\alpha$ & $0$ & $0$ &$0$ &$0$ &$0$ &$0$ \\
\hline
$S$ & $0$ & $0$ &$0$ &$0$ &$0$ &$0$\\
\hline
$y^\alpha$ & $0$ &$0$ &$0$ &$0$ &$2$ &$2$ \\
\hline
$\bar y^\alpha$ & $0$ &$0$ &$0$ &$0$ &$-2$ &$-2$ \\
\hline
$\lambda^{S\,A}$ & $1/2$ &$0$ &$1/2$ &$0$ &$-1$ &$-1$ \\
\hline
$\lambda^{S^*}_{A}$ & $0$ &$1/2$ &$1/2$ &$0$ &$1$ &$1$ \\
\hline
$\lambda^{\alpha\,A}$ & $1/2$ &$0$ &$1/2$ &$0$ &$1$ &$1$ \\
\hline
$\lambda^{\alpha^*}_{A}$ & $0$ &$1/2$ &$1/2$ &$0$ &$-1$ &$-1$ \\
\hline
$u^{A~~t}_{~\bar A}$ & $0$ & $0$ & $1/2$ &$1/2$ &$0$ &$0 $ \\
\hline
$\zeta^{\bar A t}$ & $1/2 $ & $0$ & $0$ &$1/2$ &$-1 $ &$-1$\\
\hline
$\zeta^{~~t}_{~\bar A}$ & $0 $ & $1/2$ & $0$ &$1/2$ &$1 $ &$1$\\
\hline
\hline
\end{tabular}
\end{center}
\label{topotable}
\end{table}
\section{The twist procedure}
\label{twisterie}
\setcounter{equation}{0}
\def\o#1#2{{#1\over#2}}
In this section we perform the topological twist--shift, following the
four steps pointed out in the introduction.
\par
Step i) is explicitly done following the procedure indicated in
\cite{topfgen_5,topftwist_1,topftwist_2}.
We extend the forms to ghost--forms, and we set
\begin{equation}
\hat d = d +s
\end{equation}
then we read the BRST variation of each field from the
rheonomic parametrization displayed
in Appendix C, selecting out the terms with the appropriate
ghost numbers. This step is a purely algorithmic one, and we do not
find convenient to write it in a fully extended form.
A simplified example of this calculation will be presented analyzing
step iii) and iv) of the twist--shift procedure, when we consider the
variations of the (topological) antighosts. These variations are the only
one we are ultimately interested, since they give the ``instanton"
conditions of our topological field theory.
The second step is immediate. We have analyzed in section 4
the gravitational extended
R-symmetry associated with all the fields of our model.
This global symmetry is utilized to redefine the ghost number
according to equation (\ref{brstshift}).
\par
Let us now consider with more detail steps iii) and iv).
The twist is obtained by redefining
the Lorentz group as in eq.s (\ref{newspingroup},\ref{newnewspingroup}).
The spin assignments of the fundamental
fields of our theory is resumed
in table 1.
Following the notations of references \cite{topftwist_2} we
classify each field, before the twist, by the expression
$\ ^{r}(L,R,I,Q)_f^g$, where $(L,R,I,Q)$ are the representation
labels for $(SU(2)_L, SU(2)_R, SU(2)_I, SU(2)_Q)$, $r$ is the
R--charge assignments and $f,g$ denote the ghost number and the form degree.
The twist procedure is summarized as follows:
\begin{eqnarray}
SU(2)_L &\rightarrow& SU(2)_L^\prime= {\rm diag}[SU(2)_L\otimes
SU(2)_Q ]\nonumber\\
SU(2)_R &\rightarrow& SU(2)_R^\prime= {\rm diag}[SU(2)_R\otimes
SU(2)_I ]\nonumber\\
U(1)_g &\rightarrow& U(1)_g^\prime ={\rm diag}
[U(1)_g \otimes U(1)_R ] \nonumber \\
\ ^{r}(L,R,I,Q)_f^g &\rightarrow& (L\otimes Q, I\otimes R)_f^{g+r} .
\label{paolo1}
\end{eqnarray}
The second fundamental ingredient is the topological shift.
As anticipated in section 2.2 this is a shift by a constant of
the $(0,0)^0_0$--field coming by applying the twist algorithm to the
right handed components of the supersymmetry ghost. Let us denote
this ghost by $c^A$, with spinorial components $c^{\dot \alpha \,A}$.
As it is immediately verified
$c^A$ has the following quantum numbers, before the twist:
\begin{equation}
\ ^{-1}(0,1/2,1/2,0)_0^1 .\label{paolo4}
\end{equation}
According to the prescription (\ref{paolo1}) we identify the
$SU(2)_R$ index $\dot \alpha$ with the $SU(2)_I$ index $A$, and we
perform the shift by writing
\begin{equation}
c^{\dot \alpha A} \rightarrow -\o{i}{2} e \epsilon^{{\dot \alpha}A}
+c^{\dot \alpha A} .\label{paolo5}
\end{equation}
In eq. (\ref{paolo5}), $e$ is the ``broker".
The broker, as
introduced in ref. \cite{topftwist_2}, is a zero--form with fermion
number  one and ghost number one. It is a formal object
which rearranges the
form number, ghost number and statistic in the correct way and it appears
only in the intermediate steps of the twist. $e^2$ has even fermion number
and even ghost number, and can be normalized to $e^2=1$.
\par
The BRST quantized topological field theory is thus defined by
the new set of fields, obtained from the untwisted ones by changing
the spins and the ghost numbers; and by the
shifted BRST charge, which is the sum of the old one plus the
shifted component of the supersymmetry charge.
In our approach we are not interested in writing down all
the twisted--shifted variation. We just point our attention to
the variations of the (topological) antighosts, namely the
fields $\psi^A, \lambda^{S\,A},\lambda_A^{\alpha^\star},
\zeta^{\bar A\,t}$
appearing in table 1. Such variations
(or some particular projections of these variations)
will define the instantons of our theory.
As anticipated, we are looking for $(0,0)^0_0$ component
of the supersymmetry ghost $c^{\dot \alpha A}$. Moreover, to
select the instanton conditions we set to zero all
the fields which have non zero ghost number.
\par
Let us firstly consider the variation of the right handed gravitino
$\psi^{\dot \alpha A}$. Following equation
(\ref{paolo1}) we find that
\begin{equation}
\psi^A \leftrightarrow \ ^{-1}(0,1/2,1/2,0)^0_1 \rightarrow
(0,0)_1^{-1} \oplus  (0,1)_1^{-1} .
\end{equation}
As a consequence, in the ``extended" ghost--form
$\psi^A=\psi^A + c^A$, the supersymmetry ghost $c^A$,
which has labels as described in eq. (\ref{paolo4}),
becomes, after the twist:
\begin{equation}
c^A \leftrightarrow
\ ^{-1}(0,1/2,1/2,0)^1_0 \rightarrow (0,0)_0^{0} \oplus  (0,1)_0^{0} .
\end{equation}
To read off the gravitational
instanton condition we have just to consider
the variation of the gravitino
along the $(0,0)_0^{0}$ component
of $c^A$, and to set to zero
all the non physical fields.
\par
Actually, we better consider the
gravitino with the field redefinition
$\psi^A \to e^{\o{{\cal K}}{4}} \psi^A$, in such a way that, in the
curvature definition, only the holomorphic component of the
K\"ahler connection appears.
Moreover, in presence of gauging, the K\"ahler
and the $SU(2)_I$ quaternionic connections are extended as
in Appendix C, i.e.
\begin{eqnarray}
\hat Q&=& Q + g A^\Lambda {\cal P}^0_\Lambda \nonumber\\
\hat \omega^{-x}&=& \omega^{-x} + g A^\Lambda
{\cal P}_{\Lambda}^{-x} .
\end{eqnarray}
It is quite immediate to verify that ${\cal P}^0_{\Lambda}$
does not give any contribution to the variation
of $\psi^A$ (at ghost number zero),
while the only contribution to $\hat \omega^{-x}$
come from the $SO(n)$ indices, i.e.
\begin{equation}
\hat \omega^{-x}= \omega^{-x} + g A^\alpha {\cal P}_{\alpha}^{-x}
\end{equation}
The twist procedure permits the
following identification $\psi^{\dot \alpha A}\to
\psi^{\dot \alpha \dot A}$, where we identify the
left handed Lorentz
index $\dot \alpha$
with the $SU(2)_I$ one $A=\dot A$.
Next, we define the following fields
(see reference \cite{topftwist_1}):
\begin{eqnarray}
\tilde \psi^{ab} &=&-e \bar \sigma^{ab}_{\dot \alpha\dot A}
\psi^{\dot \alpha \dot C} \epsilon_{\dot C \dot A}\\
\tilde \psi &=&-e \psi^{\dot \alpha \dot C}
\epsilon_{\dot C \dot A} \delta_{\dot A}^{\dot \alpha}
\label{paolo7}
\end{eqnarray}
where $\bar \sigma^{ab}$ are defined in appendix C
[actually here we use the euclidean version of the matrices
defined in (\ref{basta})].
Looking at the curvature definition (\ref{3.6c}) and at the
rheonomic parametrization (\ref{3.10b}) we find that the only contributions
coming from the ghost zero sector along the (shifted part)
of the supersymmetry ghost are:
\begin{eqnarray}
\delta \tilde\psi^{ab} &=& \o{i}{2}
(\omega^{-ab } -\sum_{u=1}^3I_{u}^{ab}
\hat \omega^{-u })\label{paolo101}\\
\delta {\tilde \psi}&=&\o{i}{2} Q_{{\rm hol}}(S) \label{paolo10}
\end{eqnarray}
where the matrices
$I_u^{ab}=- \o{i}{2}Tr (\bar \sigma^{ab} \sigma_u^T)$, $u=1,2,3
\equiv x,y,z$
can be identified (up to a trivial SO(3) rotation)
with the anti-selfdual
matrices $J^{-ab}_u$ introduced in (\ref{ide2},\ref{ide3}).
Eq. (\ref{paolo101})
becomes precisely the first of eq.s
(\ref{istantonequazioni}), once expressed in terms of the
curvatures.
\par
Moreover, in eq. (\ref{paolo10}), $Q_{hol}$ is given by
\begin{equation}
Q_{{\rm Hol}}=-\o{1}{4} \partial_S {\cal K} V_\mu^a \partial_a S .
\end{equation}
Therefore the instanton condition $\delta \tilde \psi=0$
corresponds, in the euclidean formalism,
to the Rey instantons. Indeed
\begin{equation}
\partial_a S = 0 \hskip 1cm \Leftrightarrow \hskip 1cm
\partial_a D =\epsilon_{abcd}\ee D H^{bcd} .
\end{equation}
\par
Let us go on and consider the instanton condition obtained
from the variation of the gaugino $\lambda^{S\,A}$.
In this case there is just a term which contribute, namely
\begin{equation}
\delta \lambda^{S\, A}=i\partial_a S \gamma^a c^A \label{paolo11}
\end{equation}
so that the instanton condition obtained from  eq. (\ref{paolo11})
is the same as the one obtained from eq. (\ref{paolo10}).
\par
Working in a similar way on
the antighost $\lambda^{\alpha^\star}_A$
and using the formul{\ae} for the metric tensor, for $G^{-\alpha^*}_{ab}$,
$Y^{\alpha^*}_{AB}$ and for $W_{AB}^{\alpha^*}$, given in appendix C,
we find the following condition
\begin{equation}\label{paolo30}
{\cal F}^{-\,\alpha\,ab}=\o{g}{2\exp{D}}
J^{-\,ab}_u {\cal P}_{\alpha}^{-\, u} .
\end{equation}
Notice that eq. (\ref{paolo30}) identify the anti self dual part of the
gauge connections with the quaternionic momentum map
${\cal P}^{-\, u}$ times the square of the {\it effective gauge coupling}.
Indeed by performing the redefinition $A^\alpha \to \o{1}{g} A^\alpha$ we
precisely get
\begin{equation}
{\cal F}^{-\,\alpha^*\,ab}= \o{1}{2}g^2_{eff.}
J^{-\,ab}_u {\cal P}_{\alpha^*}^{-\, u}
\end{equation}
with $g_{eff.}=\o{g}{\sqrt{\exp{D}}}$.
\par
Finally, the instanton condition arising from the
topological variation of the hyperini $\zeta^{\bar A t}_{\alpha}$
gives the following equations:
\begin{eqnarray}
V^{\mu [a} u_I^{b]t}\nabla_\mu q^I &=&0\nonumber\\
V^\mu_a u_I^{at} \nabla_\mu q^I &=&0 \label{paolo15}
\end{eqnarray}
where $u^{a t}$ is the vielbein defined in eq. (\ref{so4m3}).
Eq.s (\ref{paolo15}) define the so called
``gauged triholomorphic maps". To rewrite them in the more compact
notation appearing in eq. (\ref{istantonequazioni})
we have to define the three almost
quaternionic structures in $M_{\rm space-time}$
and  $HQ(m)$, namely
\begin{eqnarray}
(j_u)^\nu_\mu &\equiv &J^{-\,ab}_{u} V_{\mu a}V_b^\nu
\nonumber\\
(J_u)^I_J &\equiv & (J_{u}^-)^{~b}_a u_{J}^{at} u_{bt}^I .
\label{paolo100}\end{eqnarray}
Using eq.s (\ref{paolo100}) we can easily rewrite eq.s (\ref{paolo15})
as in eq. (\ref{istantonequazioni}) \cite{topftwist_2}
\section{Dual description of the effective theory of N=2 heterotic
string}
\setcounter{equation}{0}
In rigid N=2 supersymmetry, in order to describe the strong coupling
regime of a non--abelian gauge theory of a group ${\cal G}$ it is
useful to consider the dual effective theory which is also an N=2
gauge theory with the following differences:
\par
i) The new gauge group ${\tilde {\cal H}}=U(1)^r$ is abelian.
\par
ii) The self--interaction of the abelian gauge multiplets is encoded
in a non flat special geometry possessing a discrete group of duality
symmetries.
\par
iii) In addition to the gauge multiplet the dual theory contains a
certain number of hypermultiplets that represent the monopoles
of the original theory. This means that ${\tilde {\cal H}}$
is actually the dual of the maximal torus ${\cal H} \subset {\cal G}$
of the original gauge group.
\par
When the rigid Yang--Mills theory is embedded in a supersymmetric
theory arising as a low energy limit of heterotic superstring, it
is natural
to associate to it a Calabi--Yau threefold \cite{cerericca}
and a dual theory, which is a type II string theory
\cite{add1,add2,add3,add4} compactified on that particular manifold.
If we consider in this dual frame eq.s
(\ref{istantonequazioni}) we see that the existence of a non--trivial
monopole background of a $U(1)$ field requires, in order to be
consistent with N=2 Susy, the existence of background hypermultiplets
that are charged with respect to the Ramond--Ramond $U(1)$
gauge fields. Since these hypermultiplets carry a Ramond charge they
must appear as solitonic excitations of a type II string propagating
on the C.Y. manifold. Evidence for the existence of such states has
been given recently in \cite{sergius_5} by studying the
behaviour of the periods around the vanishing cycles of the Calabi--Yau
manifold.
\vskip 0.3truecm
\noindent
{\bf Acknowledgments}: We thank D. Anselmi, M. Bianchi, F. Fucito and
G. Rossi for useful discussions.
\newpage
\appendix
\section*{Appendix A: Structure and  parametrization of the $HQ(m)$
quaternionic manifolds}
\label{so4}
\setcounter{equation}{0}
\setcounter{section}{1}
It is possible to describe the $SO(4,m)/SO(4)\times SO(m)$ manifold as
a ``quaternionic quotient'' of the (quaternionic) projective plane $\hp$
with respect to an $SU(2)$ action. Such a description allows an explicit
parametrization of the manifold in terms of a set of quaternionic
coordinates. In the following we give such a parametrization together
with some properties of quaternionic manifolds. We have no claim to
mathematical completeness, and we refer the reader  to
\cite{momentmap_1}
for more details on the subject \par
\par
First of all, we realize the quaternionic units $e_x$, $x=1,2,3$,
satisfying the quaternionic algebra
\begin{equation}
\label{quatalgebra}
e_x e_y = -\delta_{xy} + \epsilon_{xyz} e_z
\end{equation}
by means of $2\times 2$ matrices, setting $e_x \equiv -\ii\sigma_x$. By
$\sigma_x$ we denote the standard Pauli matrices. The $e_x$ are imaginary
units since ${\bar e}_x \equiv e^{\dagger}_x = -e_x$. It will be convenient
to treat also the unit matrix on the same footing, setting
$e_0\equiv \bfone $
and thus having $\{e_a\} \equiv \{\bfone,-\ii\sigma_x\}$, $a=0,1,2,3$.
Then it is immediate to write the one-to-one correspondence
between points $\{x^a\}$ in $\IR^4$ and
quaternions $q$ by setting
\begin{equation}
q=x^a e_a = \left(\begin{array}{cc} u & \ii {\bar v} \\ \ii v & {\bar u}
                   \end{array}\right) ,
\hskip 1cm
{\bar q}=x^a {\bar e}_a = \left(\begin{array}{cc} {\bar u} & -\ii {\bar v}
                                \\ -\ii v & u \end{array}\right)
\end{equation}
where $u=x^0 -\ii x^3 $ and $v=-(x^1 +\ii x^2)$.
The quaternionic projective space $\hp$ can be described by the set of
quaternions $\{q^I\}$, $I=0,1,\ldots m+3$ satisfying
\begin{equation}
\label{quatprojdef}
\left\{\begin{array}{lll}
{\bar q}^I q^J \eta_{IJ}=\bfone & \mbox{where} &
\eta_{IJ}={\rm diag} (1,1,1,1,-1,-1,\ldots)\\
\null &\null & \null \\
\{q^I\} \sim \{q^I \hskip 2pt \nu\} & \mbox{with} &
{\bar \nu}\nu = \bfone
\end{array}\right.
\end{equation}
In eq. (\ref{quatprojdef}) the unit quaternion
$\nu$ is, in our $2\times 2$ realization, a SU(2) matrix.
\par
The above description is the
analogue of the usual description of a $\IC\IP^{N}$ space,
where the role of
the SU(2) element $\nu$ is played by a phase, i.e. an element of U(1).
Notice, however that the quaternionic product is non-commutative
and the choice of $\nu$ acting from the right in eq. (\ref{quatprojdef})
is relevant.
\par
The fundamental quaternionic one-form gauging this right SU(2) action is
\begin{equation}
\label{hpomegameno}
\omem = {\bar q}^I \dop q_I.
\end{equation}
The index are contracted with $\eta_{IJ}$; the choice of the notation $\omem$
for the $SU(2)$ connection
will be clear in the sequel. Its curvature, defined as $\Omem = \dop \omem -
\omem\wedge\omem$, is
\begin{equation}
\label{quatstruc}
\Omem = \dop{\bar q}^I \wedge \dop q_I - {\bar q}^I \dop q_I
\wedge {\bar q}^J \dop q_J.
\end{equation}
It is immediate to verify that
$\Omem$ is covariantly closed.
This 2-form is the quaternionic analogue
of the K\"ahler form of $\IC\IP^N$.
Indeed, writing $\Omem = \coeff12\sum_{x=1}^3\Omemind x e_x^T$,
we have that $\Omemind 3$
is the K\"ahler form, the metric being
\begin{equation}
\label{hpmetric}
\dop s^2 \bfone  = \dop{\bar q}^I \otimes \dop q_I -
{\bar q}^I \dop q_I \otimes {\bar q}^J \dop { q}_J
\end{equation}
Consider now the {\it left} action of an $SU(2)$ on $\hp$:
$q^I \rightarrow \mu q^I$,
with ${\bar \mu}\mu=\bfone $.
The infinitesimal action is
\begin{equation}
\label{infsu2}
\delta_x q^I = e_x q^I
\end{equation}
Such transformations leave the metric invariant, and they leave the
quaternionic structure invariant up to a gauge transformation.
This property can be reexpressed as
\begin{equation}
\label{pdefin}
{\bf i}_x \Omem = -\nabla {\cal P}^-_x,
\hskip 1cm \mbox{where} \hskip 1cm {\cal P}^-_x \propto {\bar q}^I e_x q_I
\end{equation}
where
${\bf i}_x$ denote the contraction along the killing
vector in the $x$ direction, ${\bf k}_x = e_x {\partial\over\partial q^I}
- {\partial\over\partial {\bar q}^I} e_x$.\par
The quaternionic functions ${\cal P}^-_x$ are the quaternionic momentum map
functions for the left SU(2) action. They are the key ingredient needed to
perform the quaternionic reduction of $\hp$ with respect to this action.
The quaternionic reduction procedure consists in the following two steps.
\begin{enumerate}
\item Restriction to the null level set of the momentum map,
\begin{equation}
\label{levelset}
\bigcap_x ({\cal P}^-_x)^{-1} (0).
\end{equation}
The dimension of the level set surface is ${\rm dim} \hp - 3 \times 3$
as for every quaternion ${\cal P}^-_x$ $x=1,2,3$ three real conditions
are imposed. The level set surface can be shown to be invariant with
respect to the action
of the group for which ${\cal P}^-_x$ are the momentum map functions.
\item
Quotient of the level-set surface eq. (\ref{levelset}) with respect to
the action of the group itself (in this case the left action of SU(2), eq.
(\ref{infsu2}) ).
\end{enumerate}
The dimension of the resulting quotient manifold,
which is usually denoted  as $\hp//SU(2)$, is
the dimension of the level set minus the dimension of SU(2), that is
\begin{equation}
\label{dimension}
{\rm dim} \hp//SU(2) = {\rm dim} \hp - 3\times 3 - 3 = 4 m;
\end{equation}
By the general properties of the quaternionic
reduction, the quotient manifold is quaternionic,
when it is equipped with the
quaternionic structure obtained by restricting that  of $\hp$
to the level set (eq. (\ref{levelset})) and projecting
it to the quotient.
The quaternionic quotient construction implies that we can
describe $\hp// SU(2)$ by parametrizing a set
of $4(m+4)$ quaternions $q^I$, $I=0,\ldots,m+3$ in terms of $4m$
independent real variables, so that the following equations holds:
\begin{equation}
\label{pareq}
\left\{
\begin{array}{l}
{\bar q}^I q_I = 1\\
{\bar q}^I e_x q_I = 0 \hskip 20 pt \forall x=1,2,3
\end{array}\right. .
\end{equation}
The first equation comes from the definition of the $\hp$space,
(eq.
(\ref{quatprojdef})), the other equations define the level set of the ${\cal
P}^-_x$ functions. We need to fix the gauge
for the left SU(2) acting  as $q^I\rightarrow \mu q^I$, but we  also
have  to recall that the coordinates $q^I$ were defined up to an
SU(2) acting on the right:
$q^I\rightarrow q^I\nu$, with ${\bar \nu}\nu = {\bar
\mu}\mu = \bfone$.\par
Let us use the following notation:
\begin{equation}
\label{quv1}
q^I = \left(\begin{array}{cc}U^I & \ii V^I\\ \ii {\bar V}^I & {\bar U}^I
            \end{array}\right).
\end{equation}
We split the index $I=0,1,\ldots,m+3$ into $a=0,1,2,3$ and
$t=4,5,\ldots,m+3$. We choose the quaternions
\begin{equation}
\label{quv2}
q^t = \left(\begin{array}{cc}u^t & \ii v^t\\ \ii {\bar v}^t & {\bar u}^t
            \end{array}\right).
\end{equation}
to represent the independent 4m real coordinates. In terms of the
$U^I,V^I$, the equations (\ref{pareq}) become
\begin{equation}
\label{quv3}
\left\{ \begin{array}{l}U^I U_I = 0\\ {\bar U}^I U_I =
         1/2\end{array}\right.
\hskip 1cm
\left\{ \begin{array}{l}V^I V_I = 0\\ {\bar V}^I V_I =
         1/2\end{array}\right.
\hskip 1cm
\left\{ \begin{array}{l}U^I V_I = 0\\ {\bar U}^I V_I =
         0\end{array}\right.
\end{equation}
Notice that for $V^I=0$ (and with $I$ assuming only
$m+2$ values )
these equations reduce   to the equations defining $SO(2,m)/SO(2)\times
SO(m)$,
in terms of the Calabi-Visentini coordinates $U^I\equiv Y^I$, and viceversa.
Therefore we expect the solution to the complete set of equations
to be similar to a pair of
 Calabi-Visentini systems suitably coupled.\par
Let us denote by  $u^2, u\cdot v,\ldots$ the scalar products (SO(m)
invariants) $u^t u^s\delta_{ts}$, $u^t v^s \delta_{ts}$, $\ldots$. A solution
to eq.s (\ref{quv3}) is
\begin{equation}
\label{quv4}
U ={1\over {\cal N}_U(u,v)}
\left(\begin{array}{c} 1/2(1 + u^2)\\\ii/2(1 - u^2)\\
A(u,v)\\ -\ii A(u,v)\\u^s\end{array}\right)
\hskip 1cm
V ={1\over {\cal N}_V(u,v)}
\left(\begin{array}{c} B(u,v)\\ +\ii B(u,v)\\ 1/2(1 + v^2)\\
\ii/2(1 - v^2)\\v^s\end{array}\right)
\end{equation}
where
\begin{equation}
\label{quv5}
\left\{\begin{array}{l}
A(u,v) = {1\over 1 - |u^2 v^2|^2} \left[u\cdot v - u^2 \bar u\cdot v
+ u^2 v^2 (\bar u\cdot\bar v - {\bar u}^2 u\cdot \bar v)\right] \\
\null\\
B(u,v) = {1\over 1 - |u^2 v^2|^2} \left[\bar u\cdot v - v^2 \bar
u\cdot \bar v
+ {\bar u}^2 v^2 (u\cdot\bar v - {\bar v}^2 u\cdot v)\right] \\
\end{array}\right.
\end{equation}
and where ${\cal N}_U(u,v), {\cal N}_V(u,v)$ are two normalization
constant
satisfying ${\cal N}_V(u,v) = {\cal N}_U(v,\bar u)$, which are
determined using the second row in the constraints (\ref{quv3}).
Notice that the $V^I$ are obtained from the $U^I$ by substituting
$u\rightarrow
v$, $v \rightarrow \bar u$.\par
The quaternionic structure and the metric of $\hp$, eq.s
(\ref{hpomegameno},\ref{quatstruc},\ref{hpmetric})
for the quotient manifold $\hp//SU(2)$ are obtained
by substituting the explicit parametrization of eq.s
(\ref{quv4},\ref{quv5}) for the quaternions $q^I$.
For
instance, the connection for the right SU(2) action becomes
\begin{equation}
\label{hpquotomegameno}
\omem = {\bar q}^I(u,v) \dop q_I(u,v) = {\bar q}^a(u,v) \dop q^a(u,v)
-\bar q (u,v)\cdot \dop q(u,v)
\end{equation}
$\bullet$\hskip 3pt{\it Biquaternionic structure}\par
\vskip 0.1cm
{}From now on we refer to $\hp//SU(2)$ and when we write $q^I$ we mean
$q^I(u,v)$. Beside the {\it right} SU(2) action pertinent to the
definition of $\hp$, in taking the quaternionic quotient we have
introduced into the game a {\it left} SU(2) action. Both these actions
are gauged by a connection 1-form, from which a curvature 2-form is
defined. This pair of  curvature 2--forms constitutes {\it a pair of
 independent
quaternionic structures} on $\hp//SU(2)$ that correspond to the
same metric. The metric is left invariant by {\it both}  SU(2)
actions and this restricts the holonomy group to $SU(2)\times
SU(2)\times SO(m)$. We name  quaternionic manifolds
with such a reduced holonomy as biquaternionic manifolds.
Here we just summarize our result
for $\hp//SU(2)$
\begin{equation}
\label{biquat1}
\begin{array}{cccc}
\null & \mbox{\sl Connection} & \mbox{\sl Curvature} &
\mbox{\sl Metric} \\
\null & \null & \null & \null\\
\mbox{right SU(2)} & \omem = {\bar q}^I\dop q_I & \Omem \equiv
 \dop\omem -
\omem\wedge\omem & \dop s^2 \bfone = \dop {\bar q}^I\otimes
 \dop q_I -\\
\null & \null & = \dop {\bar q}^I\wedge \dop q_I - {\bar q}^I \dop
q_I\wedge {\bar q}^J \dop q_J & - {\bar q}^I \dop
q_I\otimes {\bar q}^J \dop q_J \\
\null & \null & \null & \null \\
\mbox{left SU(2)} & \omep = \dop q^I{\bar q}_I & \Omep \equiv \dop\omep -
\omep\wedge\omep & \null\\
\null & \null & = \dop { q}^I\wedge \dop {\bar q}_I - \dop
q^I{\bar q}_I \wedge \dop q^J {\bar q}_J  &\null \\
\end{array}
\end{equation}
The "gauge'' SU(2) groups act as follows:
\begin{equation}
\label{biquat2}
\begin{array}{l}
\mbox{right SU(2)}\\
\null\\
q^I\rightarrow q^I \nu\\
\omem \rightarrow \bar\nu \omem \nu + \bar \nu \dop \nu\\
\omep \rightarrow \omep\\
\dop s^2 \rightarrow \dop s^2
\end{array}
\hskip 2cm
\begin{array}{l}
\mbox{left SU(2)}\\
\null \\
q^I\rightarrow \mu q^I \\
\omem \rightarrow \omem \\
\omep \rightarrow \mu\omep\bar\mu  +  \dop  \mu \bar \mu \\
\dop s^2 \rightarrow \dop s^2
\end{array}
\end{equation}
$\bullet$\hskip 3pt{\it The coset space $SO(4,m)/SO(4)\times SO(m)$}\par
\vskip 0.1cm
A $SO(4,m)$ matrix
$L^I_{\hskip 3pt J}$ satisfies
\begin{equation}
\label{so4m1}
L^T\eta L = \eta\hskip 1cm \mbox{i.e.}\hskip 1cm
L^I_{\hskip 3pt K} L^J_{\hskip 3pt M} \eta_{IJ} = \eta_{KM}
\end{equation}
The left-invariant 1-form $u = L^{-1} \dop L$ satisfies the Maurer-Cartan
equation $\dop u + u\wedge u=0$, that encodes the structure constants
of the algebra.
Let now $L$ be
an element of the quotient $SO(4,m)/SO(4)\times SO(m)$,
then the 1-form $u$ can be interpreted in the following way
\begin{equation}
\label{so4m3}
u = \left(\begin{array}{cc}u^{ab} & u^{at}\\ u^{ta} & u^{st}
\end{array}\right) \hskip 1cm \left\{\begin{array}{ll}
u^{ab} & \mbox{SO(4) connection} \\
u^{at} & \mbox{Vielbein on the coset}\\
u^{st} & \mbox{SO(m) connection.} \end{array}\right.
\end{equation}
Moreover the Maurer-Cartan equation can be accordingly splitted
in three equations:
\begin{equation}
\label{so4m4}
\left\{\begin{array}{ll}
\dop u^{at} +u^{ab}\wedge u^{bt} - u^{ts}\wedge u^{as} = 0 &
\mbox{Torsion equation}\\
\dop u^{ab} + u^{ac}\wedge u^{cb} =- u^{as}\wedge u^{bs}  &
\mbox{SO(4) curvature} \\
\dop u^{ts} -u^{tr}\wedge u^{rs} = u^{at}\wedge u^{as} = 0 &
\mbox{SO(m) curvature}
\end{array}\right.
\end{equation}
The above equations describe the geometry of the coset space
$\o{SO(4,m)}{SO(4)\times SO(m)}$ in terms of coset representatives.
Notice that the vielbein $u^{at}=u_I^{at} dq^I$
explicitly carries a vector index
$a=0,1,2,3$ of SO(4) and an index $t$ in the vector representation
of SO(m), which
means that the holonomy group is $SO(4)\times SO(m)$.
\vskip 0.1cm\noindent
$\bullet${\it Identification of $\hp//SU(2)$ with $SO(4,m)/SO(4)\times
SO(m)$}\par
\vskip 0.1cm
 In the above notation the identification is provided by the position
\begin{equation}
\label{ide1}
q^I={1\over 2} L^I_{\hskip 3pt a} e_a.
\end{equation}
With this position, one can easily check that the constraints eq.
(\ref{pareq}) turn into the orthogonality condition $L^I_{\hskip 3pt a}
L^J_{\hskip 3pt b} \eta_{IJ} = \delta_{ab}$.\par
In eq. (\ref{ide1}) we have converted SO(4) vectors into quaternions,
that is object transforming in the fundamental of SU(2)$\times$SU(2),
by contracting them with the imaginary units $\{e_a\}$. To show the equivalence
at
the level of the connections and curvatures we must convert the adjoint
indices of SO(4) into adjoint indices of SU(2)$\times$SU(2). This
conversion is realized by two set of $4\times 4$ antisymmetric matrices
$\{\Jpind x\}$,$\{\Jmind x\}$, $x=1,2,3$, satisfying ($\epsilon_{0123}=1$)
\begin{eqnarray}
\label{ide2}
\Jpmind x \Jpmind y &=& -\delta_{xy} +\epsilon_{xyz}\Jpmind z\nonumber\\
\Jpmind{x}_{ab} & = & \pm {1\over 2} \epsilon_{abcd} \Jpmind{x}_{cd}\nonumber\\
\left[\Jpmind x , \Jmpind y \right] & = & 0 \hskip 12pt\forall x,y.
\end{eqnarray}
They can be expressed in terms of the quaternionic units by
the following key relation:
\begin{equation}
\label{ide3}
\left\{\begin{array}{l}
\Jpind{x}_{ab} = 1/2 \hskip 3pt\trace (e_a {\bar e}_b e_x^T)\\
\Jmind{x}_{ab} = -1/2 \hskip 3pt\trace (e_a e_x^T{\bar e}_b )
\end{array}\right.
\end{equation}
The identification between the SO(4) connection $\mu^{ab}$ of
$SO(4,m)/SO(4)\times SO(m)$ and the SU(2)$\times $SU(2) connections
$\omepm$ goes as follows. Set
\begin{equation}
\label{ide4}
\omepm = {1\over 2}\omepmind x e^T_x.
\end{equation}
Then
\begin{equation}
\label{ide5}
u^{ab} = {1\over 2} (\Jpind{ab}_x \omepind x + \Jmind{ab}_x \omemind x)
\hskip 0.5cm \Leftrightarrow \hskip 0.5cm
\left\{\begin{array}{l}\omepind x= {1\over 2} \Jpind{x}_{ab} u^{ab}\\
\null\\
\omemind{x}= {1\over 2} \Jmind{x}_{ab} u^{ab}\end{array}\right. .
\end{equation}
This can be checked substituting into the explicit expressions (\ref{biquat1})
of   $\omepm$ the identification  (\ref{ide1})
of the quaternions $q^I$.\par
At the level of curvatures we analogously  set
\begin{equation}
\label{ide6}
\Omepm = \coeff12\Omepmind x e_x,
\end{equation}
and, recalling that by eq. (\ref{so4m4}) the SO(4) curvature is
$-u^{as}\wedge u^{bs}$, we have
\begin{equation}
\label{ide7}
u^{as}\wedge u^{bs} = -{1\over 2} (\Jpind{ab}_x \Omepind x +
\Jmind{ab}_x \Omemind x)
\hskip 0.5cm \Leftrightarrow \hskip 0.5cm
\left\{\begin{array}{l}\Omepind x= -{1\over 2} \Jpind{x}_{ab}
u^{as}\wedge u^{bs}\\\null\\
\Omemind x= -{1\over 2} \Jmind{x}_{ab} u^{as}\wedge u^{bs}
\end{array}\right. .
\end{equation}
Note that upon use of the definitions  (\ref{ide4},\ref{ide6})
the curvature definition $\Omepm = \dop\omepm -\omepm \wedge \omepm$ is
rewritten as $\Omepmind x = \dop\omepmind x + {1\over 2} \epsilon_{xyz}
\omepmind y \wedge \omepmind z$.
\section*{Appendix B: A note on Q--symmetry}
\label{appendiceB}
\setcounter{equation}{0}
\addtocounter{section}{1}
In order to redefine the Lorentz
group for the twist, we have to write the quaternionic vielbein as
a doublet under both the $SU(2)_I$ and $SU(2)_Q$ groups,
and as a vector under $SO(m)$.
The group $SU(2)_Q$ for the classical manifolds is the normaliser of
$SO(m)$ in the $Sp(2m)$ subgroup of the $Hol ({\cal Q M}_{4m})$.
Now, in those quantum cases, where the hypermultiplet
metric receives corrections (type II string, for instance)
it suffices that only a discrete subgroup of
$SU(2)_Q$ survives, namely it is not necessary for the vierbein
to be a doublet under a full $SU(2)_Q$ group. It is sufficient that it
is a doublet under the isometries generated by a Kleinian finite  group $G_Q$,
whose
normalizer in the holonomy group should be $SO(m)$. We name
such group the Q--symmetry group.
An interesting example is provided by the case where for $G_Q$ we take
the binary extension
of the dihedral group $D_2$. In this example
the vielbein  is acted on
by a second set of quaternionic structures (such as the $J^+_u$ we have
defined for the classical case) acting on the index $\bar A$ in
the fundamental representation of $SU(2)$. This means that
the Q--symmetry group is composed of eight elements, namely
the second set of quaternionic structures
$J^+_x, J^+_y, J^+_z$, their opposite $-J^+_x,- J^+_y,- J^+_z$
and the two matrices $\pm 1$.
This, however, is just one possibility. In the same way as any cyclic group
$\ZZ_p$ can emerge as R-symmetry group of the quantum
special manifold, in the same way any Kleinian
subgroup of $SU(2)$ can emerge as Q--symmetry of the quantum
quaternionic manifold.
\section*{Appendix C: Rheonomic parametrizations of N=2
matter coupled supergravity}
\label{appendiceC}
\setcounter{equation}{0}
\addtocounter{section}{1}
In this appendix we write the full set of rheonomic parametrization
for the matter coupled N=2 supergravity pertaining
the examples studied in this paper. These are essential ingredients
while studying the topological variation of the fields, and we include them
for completeness. Here we limit our exposition only to the essential
points and to the formulae that are needed in the present paper.
For a detailed treatment on this subject we
refer to \cite{skgsugra_1}.
To write the set of curvature definitions and rheonomic parametrization
we need to recall a procedure named in \cite{skgsugra_1}
``gauging of the composite connection".
On the scalar manifold
$ST (n) \times HQ (m)$
we can introduce several connection 1-forms related to different bundles.
In particular we have the  standard Levi--Civita connection
and the $SU(2) \times U(1)$ connection $(\omega^- , {\cal Q})$,
as defined in
(\ref{quatstruc}) and (\ref{parapiglia}).
Gauging the corresponding supergravity
theory is done by gauging these composite connections in the
underlying $\sigma$-model.
For a K\"ahler manifold, if we call $z^i$
the scalar fields\footnote{For
the manifolds $ST(n)$ considered in the present paper
we have $z^i=\{z^0,z^\alpha\}=S,Y^\alpha$, $\alpha=1,\ldots n$}
and $k^i (z)$ the Killing vectors, we have to replace the ordinary
differential by the covariant ones:
\begin{equation}
\label{reo1}
dz^i \rightarrow \nabla z^i=dz^i + g A^\Lambda k_\Lambda (z)
\end{equation}
together with their complex conjugate.
In eq. (\ref{reo1}) $A^\Lambda$ is the gauge one form
($\Lambda=0,S,\alpha$ in our case).
At the same time the Levi--Civita connection
$\Gamma^i_j=\Gamma^i_{jk}dz^k$ is replaced by:
\begin{equation}
\Gamma^i_j \rightarrow \hat \Gamma^i_{j}\equiv
\Gamma^i_{jk} \nabla z^k +g A^\Lambda \partial_j k_\Lambda^i
\end{equation}
so that the curvature two form become
(as in the previous equations we omit the obvious
complex conjugate expression)
\begin{equation}
\hat R^i_j= R_{jkl}^i \nabla z^k \wedge \nabla z^l + g
{\cal F}^\Lambda \partial_j k^i_\Lambda
\end{equation}
where ${\cal F}$ is the field strength associated with
$A^\Lambda$.
In a fully analogous way we can gauge the $Sp(2m)$ connection
of the quaternionic scalar manifold, but we will
now focus our attention on the
$SU(2) \times U(1)$ connection. In this case the existence
of the Killing vector prepotentials
${\cal P}^0_\Lambda,{\cal P}^{-x}_\Lambda$ ($x=1,2,3$) permits
the following covariant definitions:
\begin{eqnarray}
{\cal Q}&\rightarrow& \hat {\cal Q}= {\cal Q} +
g A^\Lambda {\cal P}^0_\Lambda
\nonumber\\
\omega^{-x} &\rightarrow &\hat \omega^{-x} =\omega^{-x} + g A^\Lambda
{\cal P}^{-x}_\Lambda
\end{eqnarray}
where ${\cal P}^{-x}_\Lambda$ is given in eq. (\ref{pdefin}) and
${\cal P}^0_\Lambda$ is defined by the relation
\begin{equation}
i_\Lambda {\cal K}= -d {\cal P}^0_\Lambda
\end{equation}
In computing the associated gauged curvatures we get:
\begin{eqnarray}
\hat {\cal K}&=&i g_{i j^*} \nabla z^i \wedge \nabla z^{j^*} +
g {\cal F}^\Lambda {\cal P}_\Lambda^0 \nonumber\\
\hat \Omega^{-x} &=& \Omega^{-x}_{IJ} \nabla q^I \wedge \nabla q^J
+ g {\cal F}^\Lambda {\cal P}_\Lambda^{-x}
\end{eqnarray}
where
\begin{equation}
\nabla q^I= d q^I +g A^\Lambda k_{\Lambda}^I (q)\label{c10}
\end{equation}
$k^I_\Lambda (q)$ being the quaternionic Killing vectors.
We are now able to write down the full set of curvature definitions
and rheonomic parametrizations of the N=2 matter coupled supergravity.
We start with the hypermultiplets in the ungauged case. In the notation
appearing in table 1 we have the positive and negative
chirality hyperini $\zeta^{\bar A t}, \zeta_{\bar A}^t$.
For the ungauged case we can write the following curvature definition
for the right handed hyperino (a similar one holds for the other):
\begin{equation}
\nabla \zeta^{\bar A t}=d \zeta^{\bar A t} - {1\over 4}
\gamma_{ab} \omega^{ab}\zeta^{\bar A t} -
\Delta_{\bar B s}^{\bar A t} \zeta^{\bar B s} +{i\over 2}
{\cal Q} \zeta^{\bar A t}
\label{c1}
\end{equation}
In the above equation $\Delta^{\bar B s \,\bar A t}$ is the
$Sp(2m)$ connection. Indeed in our example the symplectic index $\alpha$ is
splitted into an  index ${\bar A}$ of $SU(2)_Q$ times an index $t$ of
$SO(m)$. The raising and lowering of the symplectic indices is realized by
\begin{equation}
C_{\alpha\beta}\equiv C_{{\bar A}t, {\bar B}s} =
\epsilon_{\bar B \bar A}\delta_{st} .
\end{equation}
Moreover in eq. (\ref{c1})
\begin{equation}
\gamma_{ab}=
\o{1}{2} [ \gamma_a, \gamma_b ]
\equiv \twomat{2\sigma_{ab}}{0}{0}{2\bar \sigma_{ab}}
\label{basta}
\end{equation}
where we choose (in Minkowskian notation)
\begin{eqnarray}
\gamma^a &=&\twomat{0}{\sigma^a}{\bar \sigma^a}{0}\nonumber\\
(\bar \sigma^a)^{\dot \alpha \alpha} &=&
\epsilon^{\dot \alpha \dot\beta}
\epsilon^{\alpha \beta} (\sigma^a)_{\beta \dot \beta}
\end{eqnarray}
with $\sigma^0=\hbox{diag} (-1,-1)$.
The superspace parametrization of
the quaternionic vielbein $u^{A~~t}_{~\bar A}$ is given by
\begin{equation}
u^{A~~t}_{~\bar A}=u^{A~~t}_{a\bar A}V^a +\epsilon^{AB}
\bar \psi_B \zeta^{\bar B  t} \epsilon_{\bar B \bar A}
+ \bar \psi^A \zeta^t_{\bar A}
\label{lassu}
\end{equation}
Eq. (\ref{lassu}) just fixes the supersymmetry transformation law
of the quaternionic coordinate $q^I$.
The rheonomic parametrization
$\nabla \zeta^{\bar A t}$
compatible with the Bianchi identity
coming from eq. (\ref{c1}) is the following one:
\begin{equation}
\label{hy1}
\nabla\zeta^{\bar A t} =\nabla_a \zeta^{\bar A t} V^a
+i u_{a~~\bar B}^{B~~~s}\gamma^a \psi^A \epsilon_{AB} \epsilon^{\bar A
\bar B} \delta_{st}
\end{equation}
For the gauged case we have just to replace the $\nabla$ derivative
appearing in (\ref{c1}), which is covariant with respect to
the spin, K\"ahler and $Sp(2m)$ connection with a derivative
$\hat \nabla$, covariant also with respect to the gauge connection.
This substitution implies the
following change in the rheonomic parametrization:
\begin{equation}
\hat \nabla \zeta^{\bar A t}= \hat \nabla \zeta^{\bar A t}_{old}
+2g u^{A~~t}_{I~\bar A} \epsilon^{\bar B \bar A}
k^I_\Lambda (q){\bar L}^\Lambda\psi_A
\end{equation}
\par
The ungauged curvature definition of the gravitational sector are:
\begin{eqnarray}
R^a &=& {\cal D}V^a-i\bar\psi_A\wedge \gamma^a\psi^A\label{3.6a}\\
\rho_A &=& d\psi_A-{1\over 4} \gamma_{ab} \omega^{ab}\wedge\psi_A+
{i\over 2} {\cal Q}\wedge \psi_A -
\omega_A^{~B}\wedge \psi_B
\equiv \nabla \psi_A \label{3.6b} \\
\rho^A &=& d\psi^A-{1\over 4} \gamma_{ab} \omega^{ab}\wedge\psi^A
-{i\over 2} {\cal Q}\wedge\psi^A
+\omega_{~A}^{B} \wedge \psi^B \equiv \nabla \psi^A \label{3.6c} \\
R^{ab} &=& d\omega^{ab}-\omega^a_c\wedge \omega^{ab}\label{3.6d}
\end{eqnarray}
where $\omega_A^{~B}=1/2 i (\sigma_x)_A^{~B} \omega^{-x}$ and
$\omega_{~A}^B=\epsilon^{AL} \omega_L^{~M} \epsilon_{MB}$.
For the vector multiplet we define, together with the
differentials  $dz^i$, $d\bar z^{i^*}$  (``curvatures" of $z^i$,
$\bar z^{i^*}$), the following superspace field strengths:
\begin{eqnarray}
\nabla\lambda^{iA} &\equiv & d\lambda^{iA}-{1\over 4} \gamma_{ab}
\omega^{ab} \lambda^{iA} -{i\over 2} {\cal Q}\lambda^{iA}+
\Gamma^i_{\phantom{i}j}\lambda^{jA}+\omega^{A}_{~B} \wedge
\lambda^{iB}
\label{3.7a}
\\
\nabla\lambda^{i^*}_A &\equiv &d\lambda^{i^*}_A-{1\over 4} \gamma_{ab}
\omega^{ab}\lambda^{i^*}_A+{i\over 2} {\cal Q}\lambda^{i^*}_A+
\Gamma^{i^*}_{\phantom{i^*}j^*}\lambda^{j*}_A -\omega_{A}^{~B} \wedge
\lambda^{i^*}_B
\label{3.7b}
\\
F^\Lambda &\equiv & dA^\Lambda+\bar L^\Lambda \bar\psi_A\wedge\psi_B
\epsilon^{AB}+L^\Lambda\bar\psi^A\wedge \psi^B\epsilon_{AB}
\label{3.7c}
\end{eqnarray}
where $\Gamma^i_j$ is the Levi--Civita connection
and $L^\Lambda=e^\cK X^\Lambda$.

The complete parametrizations of the curvatures,
consistent with Bianchi identities
following from eq.s  (\ref{3.6a})--(\ref{3.7c}), are given by
\begin{eqnarray}
R^a &=& 0 \label{3.9}\\
\rho_A &=&\rho_{A\vert ab} V^a\wedge V^b+\{(A_{A}^{\ \vert b B}
\eta_{ab}+A_{A}^{\prime \vert b\ B}\gamma_{ab})\psi_B+ \nonumber\\
&+&( \epsilon_{AB} T^+_{ab})
\gamma^b\psi^B\}\wedge V^a
-\o{\ii}{4}\epsilon_{AB}\epsilon^{{\bar A}{\bar B}}
{\bar\zeta}^{t}_{\bar A} \gamma_{ab}\zeta^{t}_{\bar B} \,\gamma^b \psi^B V^a
\label{3.10a}\\
\rho^A &=& \rho^A_{\vert ab} V^a \wedge V^b+
\{ (\bar A^{A \vert b}_{~B} \eta_{ab}+
\bar A^{\prime A\vert b}_{B} \gamma^{ab})\psi_B \nonumber\\
& + &(\epsilon^{AB} T_{ab}^{-})\gamma^b\psi_B\} \wedge V^a
-\o{\ii}{4}\epsilon^{AB}\epsilon_{{\bar A}{\bar B}}
{\bar\zeta}^{{\bar A}t} \gamma_{ab}\zeta^{{\bar B}t} \,\gamma^b \psi_B V^a
\label{3.10b}
\\
R^{ab} &=& R^{ab}_{\phantom{ab}cd} V^c\wedge V^d-i
(\bar\psi_A\theta^{A\vert ab}_c +\bar\psi^A \theta^{ab}_{A\vert c})
\wedge V^c+\epsilon^{abcf} \bar\psi^A\wedge \gamma_f\psi_B
(A^{\prime B}_{\phantom{B}A\vert c}-\bar A^{\prime B}_{\phantom{B}A\vert c})
\nonumber\\
&+& i\epsilon^{AB}\bar\psi_A\wedge
\psi_B T^{+ab} -i \epsilon_{AB} \bar\psi^A\wedge\psi^B T^{-ab}\label{3.11}
\end{eqnarray}
\begin{eqnarray}
F^\Lambda &=& F^\Lambda_{ab} V^a\wedge V^b+(if^\Lambda_i\bar\lambda^{iA}
\gamma^a\psi^B\epsilon_{AB} +i f^\Lambda_{i^*} \bar\lambda^{i^*}_A
\gamma^a \psi_B \epsilon^{AB})\wedge V_a \label{3.12}\\
\nabla\lambda^{iA} &=&\nabla_a\lambda^{iA} V^a+iZ^i_a\gamma^a\psi^A+
G^{+i}_{ab} \gamma^{ab} \psi_B \epsilon^{AB}+ Y^{iAB}\psi_B
\label{3.13a}\\
\nabla\lambda^{i^*}_A &=& \nabla_a \lambda^{i^*}_A V^a+i \bar Z^{i^*}_a
\gamma^a\psi_A+G^{-{i^*}}_{ab} \gamma^{ab} \psi^B \epsilon_{AB}
+ Y^{i^*}_{\phantom{i^*}AB}\psi^B\label{3.13b}\\
dz^i &=& Z^i_a V^a+\bar\lambda^{iA}\psi_A\label{3.14a} \\
d\bar z^{i^*} &=& Z^{i^*}_a V^a+\bar\lambda^{i^*}_A\psi^A \label{3.14b}
\end{eqnarray}
where
\begin{eqnarray}
A_{A\ \vert a}^{\phantom{A\ \vert a}B} &=&-{i\over 4} g_{k^*\ell}
(\bar\lambda^{k^*}_A\gamma_a \lambda^{\ell B}-\delta^B_A
\bar\lambda^{k^*}_C \gamma_a\lambda^{\ell C})\label{3.16a}\\
A^{\prime B}_{A\ \vert a} &=&{i\over 4} g_{k^*\ell}(\bar\lambda^{k^*}_A\gamma_a
\lambda^{\ell B}-{1\over 2} \delta^B_A \bar\lambda^{k^*}_C
\gamma_a\lambda^{C\ell})+{i\over 4} \delta_A^B \bar \zeta_{\bar A}^t
\gamma_a \zeta^{\bar A t}\label{3.16b}
\end{eqnarray}
\begin{equation}
S_{AB}=\bar S^{AB}=0\label{3.17}
\end{equation}
\begin{equation}
\theta^{ab}_{cA}=2\gamma^{[a}\rho^{b]c}_A+\gamma^c\rho^{ab}_A;
\quad
\theta^{ab\ A}_c=2\gamma^{[a}\rho^{b]c\vert A}+\gamma^c
\rho^{ab\vert A}\label{3.18}
\end{equation}
\begin{eqnarray}
T^+_{ab} &=&
2\ii {\rm Im}\cN_{\Lambda\Sigma} L^\Sigma
({\cal F}_{ab}^{\Lambda +} +{1\over 8} \nabla_i f^\Lambda_j
\bar \lambda^{i A} \gamma_{ab} \lambda^{jB} \epsilon_{AB}
-\o 14 \epsilon_{{\bar A}{\bar B}}{\bar\zeta}^{{\bar A}t}\gamma_{ab}
\zeta^{{\bar B}t}\, L^\Lambda
)\nonumber\\
T^-_{ab} &=&
2\ii {\rm Im}\cN_{\Lambda\Sigma} {\bar L}^\Sigma
({\cal F}_{ab}^{\Lambda -} +{1\over 8} \nabla_{i^*} \bar
f^\Lambda_{j^*}
\bar \lambda^{i^*}_A \gamma_{ab} \lambda^{j^*}_B \epsilon^{AB}
-\o 14 \epsilon^{{\bar A}{\bar B}}{\bar\zeta}_{\bar A}^t\gamma_{ab}
\zeta_{\bar B}^t\, {\bar L}^\Lambda
)
\label{c20}
\end{eqnarray}
\begin{eqnarray}
G^{i+}_{ab} &=& -g^{i^*j} \bar f^\Gamma_{j^*}
{\rm Im}{\cal N}_{\Gamma\Lambda}
\left ( {\cal F}^{\Lambda +}_{ab} +{1\over 8}
\nabla_{i}  f^{\Lambda}_{j} \bar \lambda^{iA}
\gamma_{ab} \lambda^{jB} \epsilon_{AB}
-\o 14 \epsilon_{{\bar A}{\bar B}}{\bar\zeta}^{{\bar A}t}\gamma_{ab}
\zeta^{{\bar B}t}\, L^\Lambda
\right )\nonumber\\
G^{i^*-}_{ab} &=& -g^{i^*j} f^\Gamma_j {\rm Im}{\cal N}_{\Gamma\Lambda}
\left ( {\cal F}^{\Lambda -}_{ab} +{1\over 8}
\nabla_{i^*} \bar f^{\Lambda}_{j^*} \bar \lambda^{i^*}_A
\gamma_{ab} \lambda^{j^*}_B \epsilon^{AB}
-\o 14 \epsilon^{{\bar A}{\bar B}}{\bar\zeta}_{\bar A}^t\gamma_{ab}
\zeta_{\bar B}^t\, {\bar L}^\Lambda
\right )\nonumber\\
\label{c30}
\end{eqnarray}
\begin{eqnarray}
Y^{ABi} &=& g^{ij^*} C_{j^*k^*\ell^*} \bar\lambda^{k^*}_C
\lambda^{\ell^*}_D
\epsilon^{AC} \epsilon^{BD}\label{3.21a}
\\
Y^{i^*}_{AB}& =& g^{i^*j} C_{ik\ell} \bar\lambda^{kC}
\lambda^{\ell D} \epsilon_{AC} \epsilon_{BD}\label{3.21b}
\end{eqnarray}
\par
The special geometry gadgets $L^\Lambda, \bar L^\Lambda,
f^\Lambda_i, f^\Lambda_{i^*}$ and the
tensors $C_{ijk}$, and $C_{{i^*}{j^*} k^*}$
turn out to be constrained by consistency
of the Bianchi identities as it
follows
\begin{equation}
\nabla_{i^*} L^\Lambda = \nabla_i \bar L^\Lambda =0\label{3.22}
\end{equation}
\begin{equation}
f^\Lambda_i=\nabla_i L^\Lambda;\quad
f^\Lambda_{i^*} = \nabla_{i^*} L^\Lambda\label{3.23}
\end{equation}
\begin{eqnarray}
\nabla_{\ell^*} C_{ijk} &=& \nabla_\ell C_{i^*j^*k^*}=0\label{3.24}\\
\nabla_{[\ell} C_{i]jk} &=& \nabla_{[\ell^*} C_{i^*]j^*k^*}=0\label{3.25}\\
\ii g^{i\ell^*} f^\Lambda_{\ell^*} C_{ijk} &=& \nabla_{j}
f^\Lambda_{k}\label{3.26}
\end{eqnarray}
We do not report the explicit calculation to prove the above equations,
 but we stress that they are fully determined by the Bianchi identities
of N=2 supergravity.
The solution for $C_{ijk}$ can be expressed by (\cite{ceretoine})
\begin{equation}
C_{ijk}= 2\ii {\rm Im}\cN_{\Lambda\Sigma}\, f^\Lambda_i
\nabla_{j} f^\Sigma_{k}
\label{3.27}\end{equation}
\par
In the gauged case we have firstly to replace in the curvature definitions
$\nabla$ with $\hat \nabla$, namely the derivative covariant with respect
to the gauge field. Secondly, the new parametrization will contain
extra terms with respect to the old ones which are proportional to the
gauge coupling constant $g$.
In particular the new parametrization are:
\begin{eqnarray}
R^a &=& 0 \label{bla0}\\
\hat \rho_A &=& \hat \rho^{(old)}_A + i g S_{AB} \gamma_a
\psi^B \wedge V^a \label{bla1}\\
\hat \rho^A &=& \hat \rho^{A\,(old)} + i g \bar S^{AB} \gamma_a
\psi_B \wedge V^a \label{bla2}\\
\hat R^{ab}&=&\hat R^{ab\, (old)}
-\bar \psi_A \wedge\gamma^{ab} \psi_B g \bar S^{AB}
-\bar \psi^A \wedge\gamma^{ab} \psi^B g S_{AB}\\
F^\Lambda &=&F^{\Lambda\, (old)}\\
\hat \nabla \lambda^{iA}&=&
\hat \lambda^{iA\, (old)}+ g W^{iAB} \psi_B\label{bla3}\\
\hat \lambda^{i^*}_A &=&
\hat \lambda^{i^*\, (old)}_A+ g W^{i^*}_{AB} \psi^B\label{bla4}\\
\hat \nabla z^i &=&\nabla z^{i\,old}
\quad \hat \nabla z^{i^*} =\nabla z^{i^*\, (old)}
\label{c40}
\end{eqnarray}
together with equation (\ref{c1}) for the hyperinos.
In eq. (\ref{c40}) $S_{AB}$ and the corresponding conjugated
expression is given by:
\begin{eqnarray}
S_{AB}&=&{1\over 2} i (\sigma_x)^{~C}_A \epsilon_{BC}
{\cal P}^x_{\Lambda}L^\Lambda \nonumber\\
\bar S^{AB}&=&{1\over 2} i (\sigma_x)_{C}^{~B} \epsilon^{CA}
{\cal P}^x_{\Lambda}\bar L^\Lambda
\end{eqnarray}
while  $W^{iAB}$ is given by the sum of a symmetric part
plus an antisymmetric one, where
\begin{eqnarray}
W^{i[AB]}&=&\epsilon^{AB}k_{\Lambda}^i \bar L^\lambda \nonumber\\
W^{i(AB)} &=& -i(\sigma_x)_C^{~B} \epsilon^{CA} {\cal P}^x_\Sigma
g^{ij^*} f_{j^*}^\Sigma
\label{c90}
\end{eqnarray}

\end{document}